\documentclass[sigconf]{acmart}
\renewcommand\footnotetextcopyrightpermission[1]{}

\usepackage{fancyhdr}
\usepackage{def}
\usepackage[ruled,vlined,linesnumbered,noresetcount]{algorithm2e}
\setcitestyle{square}
\usepackage{comment}
\usepackage{graphicx}
\usepackage{subcaption}
\usepackage{booktabs}
\usepackage{xspace}
\usepackage{amsmath}
\usepackage{mathtools}
\usepackage{stfloats}
\usepackage{booktabs}
\usepackage{multirow}

\usepackage{algpseudocode}% http://ctan.org/pkg/algorithmicx
\usepackage[toc]{appendix}

\usepackage{listings}
\definecolor{dkgreen}{rgb}{0,0.6,0}
\definecolor{gray}{rgb}{0.5,0.5,0.5}
\definecolor{mauve}{rgb}{0.58,0,0.82}

\newcommand{\cxlmem}{CXL memory\xspace}
\newcommand{\cxlio}{\texttt{CXL.io}\xspace}
\newcommand{\cxlcache}{\texttt{CXL.cache}\xspace}

\algnewcommand{\algorithmicgoto}{\textbf{go to}}%
\algnewcommand{\Goto}{\algorithmicgoto\xspace}%
\algnewcommand{\Label}{\State\unskip}

\newcommand\Tstrut{\rule{0pt}{2.4ex}}       	% "top" strut
\newcommand\Bstrut{\rule[-1.0ex]{0pt}{0pt}} 	% "bottom" strut
\newcommand{\TBstrut}{\Tstrut\Bstrut}			 % top&bottom strut

\def\arch{\texttt{memo}\xspace}
\def\policy{\texttt{Caption}\xspace}

\AtBeginDocument{%
  }

\begin{document}
\title{Demystifying CXL Memory with \\ Genuine CXL-Ready Systems and Devices}

\settopmatter{authorsperrow=4}
\author{Yan Sun}
\affiliation{%
  \institution{University of Illinois}
  \city{Urbana}
  \country{U.S.A.}}
\email{yans3@illinois.edu}
\author{Yifan Yuan}
\affiliation{%
  \institution{Intel Labs}
  \city{Hillsboro}
  \country{U.S.A.}}
\email{yifan.yuan@intel.com}
\author{Zeduo Yu}
\affiliation{%
  \institution{University of Illinois}
  \city{Urbana}
  \country{U.S.A.}}
\email{zeduoyu2@illinois.edu}
\author{Reese Kuper}
\affiliation{%
  \institution{University of Illinois}
  \city{Urbana}
  \country{U.S.A.}}
\email{rkuper2@illinois.edu}
\author{Chihun Song}
\affiliation{%
  \institution{University of Illinois}
  \city{Urbana}
  \country{U.S.A.}}
\email{chihuns2@illinois.edu}
\author{Jinghan Huang}
\affiliation{%https://www.overleaf.com/project/63c18849d9e63dbee93dc074
  \institution{University of Illinois}
  \city{Urbana}
  \country{U.S.A.}}
\email{jinghan4@illinois.edu}
\author{Houxiang Ji}
\affiliation{%
  \institution{University of Illinois}
  \city{Urbana}
  \country{U.S.A.}}
\email{hj14@illinois.edu}
\author{Siddharth Agarwal}
\affiliation{%
  \institution{University of Illinois}
  \city{Urbana}
  \country{U.S.A.}}
\email{sa10@illinois.edu}
\author{Jiaqi Lou}
\affiliation{%
  \institution{University of Illinois}
  \city{Urbana}
  \country{U.S.A.}}
\email{jiaqil6@illinois.edu}
\author{Ipoom Jeong}
\affiliation{%
  \institution{University of Illinois}
  \city{Urbana}
  \country{U.S.A.}}
\email{ipoom@illinois.edu}
\author{Ren Wang}
\affiliation{%
  \institution{Intel Labs}
  \city{Hillsboro}
  \country{U.S.A.}}
\email{ren.wang@intel.com}
\author{Jung Ho Ahn}
\affiliation{%
  \institution{Seoul National University}
  \city{Seoul}
  \country{Republic of Korea}}
\email{gajh@snu.ac.kr}
\author{Tianyin Xu}
\affiliation{%
  \institution{University of Illinois}
  \city{Urbana}
  \country{U.S.A.}}
\email{tyxu@illinois.edu}
\author{Nam Sung Kim}
\affiliation{%
  \institution{University of Illinois}
  \city{Urbana}
  \country{U.S.A.}}
\email{nskim@illinois.edu}

\renewcommand{\shortauthors}{Sun et al.}

\newcommand{\yan}[1]{{\color{brown}[\textbf{\sc yan}: \textit{#1}]}}

% !TEX root = paper.tex
\begin{abstract}

\footnote{This work has been accepted by a conference. The authoritative version of this work  will appear in the Proceedings of the IEEE/ACM International Symposium on Microarchitecture (MICRO), 2023. Please refer to \url{https://doi.org/10.1145/3613424.3614256} for the official version of this paper.}The ever-growing demands for memory with larger capacity and higher bandwidth %by datacenter servers
have driven recent innovations on memory expansion and disaggregation technologies based on Compute eXpress Link %\textsuperscript{TM} (CXL\textsuperscript{TM}).
(CXL).
Especially, CXL-based memory expansion technology has recently gained notable attention for its ability not only to economically expand memory capacity and bandwidth but also to decouple memory technologies from a specific memory interface of the CPU.
However, since \cxlmem devices %and systems that support them 
have not been widely available,
%due to the limited availability of \cxlmem devices, 
they have been emulated using DDR memory in a remote NUMA node. %a remote NUMA node in a multi-socket system.
%
%Recent studies on the performance characterizations and efficient uses of CXL memory commonly treat CXL-memory systems as conventional NUMA systems, where CXL memory is emulated using a remote NUMA node %(without CPU cores and caches) 
%in a multi-socket system. 
%
In this paper, for the first time, we comprehensively evaluate a true CXL-ready system based on the latest 4\textsuperscript{th}-generation Intel Xeon  CPU 
%(Sapphire Rapids) 
with three CXL memory devices from different manufacturers. 
Specifically, we run a set of microbenchmarks not only to compare the performance of true CXL memory with that of emulated CXL memory %(\ie, DDR memory in a remote NUMA node) 
but also to analyze the complex interplay between the CPU and CXL memory in depth. 
This reveals important differences between emulated \cxlmem and true \cxlmem, some of which will compel researchers to revisit the analyses and proposals from recent work.
Next, we identify opportunities for memory-bandwidth-intensive applications to benefit from the use of \cxlmem.
Lastly, we propose a CXL-memory-aware dynamic page allocation policy, \policy to more efficiently use CXL memory as a bandwidth expander.
We demonstrate that \policy can automatically converge to an empirically favorable percentage of pages allocated to \cxlmem, which improves the performance of memory-bandwidth-intensive applications by up to 24\% when compared to the default page allocation policy designed for traditional NUMA systems. 

\end{abstract}


% \ccsdesc[500]{Hardware~Memory and dense storage}
% \ccsdesc[500]{Computer systems organization~Architectures}
% \ccsdesc[500]{General and reference~Measurement}


\maketitle
\pagestyle{plain}

\section{Introduction}
\label{sec:intro}
Emerging applications have demanded memory with even larger capacity and higher bandwidth at lower power consumption. 
However, as the current memory technologies have almost reached their scaling limits, it has become more challenging to meet these demands cost-efficiently. 
Especially, when focusing on the memory interface technology, we observe that DDR5 requires 288 pins per channel~\cite{jedec}, making it more expensive to increase the number of channels for higher bandwidth under the CPU's package pin constraint. 
Besides, various signaling challenges in high-speed parallel interfaces, such as DDR, make it harder to further increase the rate of data transfers.
This results in super-linearly increasing energy consumption per bit transfer~\cite{8326997} and reducing the number of memory modules (DIMMs) per channel to one for the maximum bandwidth~\cite{4thGenIn27:online}. 
As the capacity and bandwidth of memory are functions of the number of channels per CPU package, the number of DIMMs per channel, and the rate of bit transfers per channel, DDR has already shown its limited bandwidth and capacity scalability. 
This calls for alternative memory interface technologies and memory subsystem architectures. 

\begin{figure}[t]
    \centering
    %\vspace{-15pt}
    \includegraphics[width=1\linewidth]{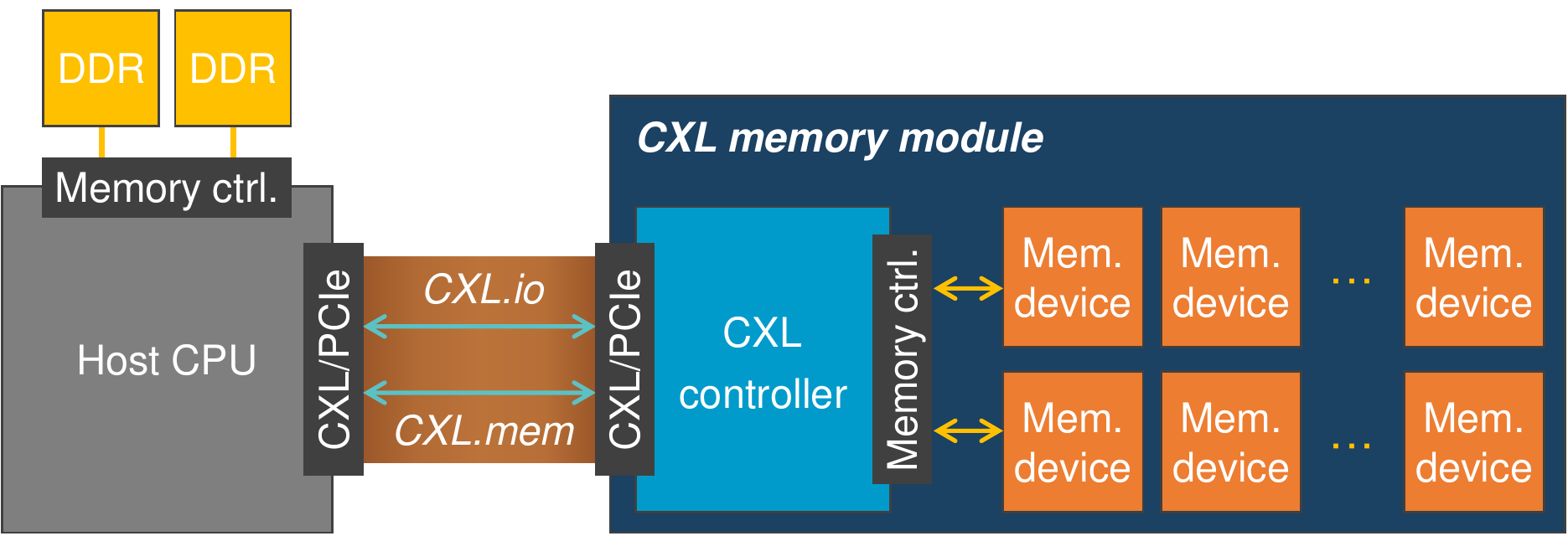}
    \vspace{-12pt}
    \caption{CXL memory module architecture.} 
    \vspace{-6pt}
    \label{fig:cxl-mem}
\end{figure}

Among them, Compute eXpress Link (CXL)~\cite{cxl2} has emerged as one of the most promising memory interface technologies. 
CXL is an open standard developed through a joint effort by major hardware manufacturers and hyperscalers. %cloud service providers. 
As CXL is built on the standard PCIe, which is a serial interface technology, it can offer much higher bit-transfer rates per pin (\eg, PCIe 4.0: 16~Gbps/lane vs. DDR4-3200: 3.2~Gbps/pin) and consumes lower energy per bit transfer (\eg, PCIe 4.0: 6~pJ/bit~\cite{LowPower73:online} vs. DDR4: 22~pJ/bit~\cite{8377983}), but at the cost of much longer link latency (\eg, PCIe 4.0: $\sim$40~ns~\cite{cxl2} vs. DDR4: $<$1~ns~\cite{6509675}).
Compared to PCIe, CXL implements additional features that enable the CPU to communicate with devices and their attached memory in a cache-coherent fashion using load and store instructions. 
\figref{fig:cxl-mem} illustrates a CXL memory device consisting of a CXL controller and memory devices.
Consuming $\sim$3$\times$ fewer pins than DDR5, a CXL memory device based on PCIe 5.0 $\times$8 may expand memory capacity and bandwidth of systems cost-efficiently.
Furthermore, with the CXL controller between the CPU and memory devices, CXL decouples memory technologies from a specific memory interface technology supported by the CPU.
This grants memory manufacturers unprecedented flexibility in designing and optimizing their memory devices.
Besides, by employing retimers and switches, a CPU with CXL support can easily access memory in remote nodes with lower latency than traditional network interface technologies like RDMA, efficiently facilitating memory disaggregation.
%not only for memory capacity/bandwidth expansion but also for memory disaggregation. % in both the industry and academia. 
%
These advantages position memory-related extension as one of the primary target use cases for CXL~\cite{micron-cxl, montage-cxl, rambus-cxl, smdk_github}, and major hardware manufacturers have announced CXL support in their product roadmaps~\cite{spr, amd_4th_gen_epic, micron-cxl, montage-cxl, smdk_github}.

Given its promising vision, \cxlmem has recently attracted significant attention with active investigation for datacenter-scale deployment~\cite{tpp,pond}. 
Unfortunately, due to the lack of commercially available hardware with CXL support, most of the recent research on \cxlmem has been based on emulation using memory in a remote NUMA node in a multi-socket system, since \cxlmem is exposed as such~\cite{pond, tpp,10.1145/3545008.3545054}. 
However, as we will reveal in this paper, there are fundamental differences between emulated \cxlmem and true \cxlmem. 
That is, the common emulation-based practice of using a remote NUMA node to explore \cxlmem may give us misleading performance characterization results and/or lead to suboptimal design decisions.

This paper addresses a pressing need to understand the capabilities and performance characteristics of true \cxlmem, as well as their impact on the performance of (co-running) applications and the design of OS policies to best use \cxlmem.
To this end, we take a system based on the CXL-ready 4\textsuperscript{th}-generation Intel Xeon  CPU~\cite{spr} and three \cxlmem devices from different manufacturers (\S\ref{sec:setup}).
Then, for the first time, we not only compare the performance of true \cxlmem %based on three different devices 
with that of emulated \cxlmem, but also conduct an in-depth analysis of the complex interplay between the CPU and \cxlmem.
Based on these comprehensive analyses, we make the following contributions. 

\niparagraph{\cxlmem $\neq$ remote NUMA memory (\S\ref{sec:character}).}  
We reveal that true \cxlmem exhibits notably different performance \textbf{\underline{C}}haracteristics  
from emulated \cxlmem.
\textbf{(C1)} Depending on CXL controller designs and/or memory technologies, true \cxlmem devices give a wide range of memory access latency and bandwidth values.
\textbf{(C2)} True \cxlmem  can give up to 26\% lower latency and 3--66\% higher bandwidth efficiency than emulated \cxlmem, depending on memory access instruction types and \cxlmem devices. 
%
%This is because true \cxlmem, without caches, does not incur the overhead of cache coherence checks in contrast to emulated \cxlmem.
This is because true \cxlmem has neither caches nor CPU cores that modify caches, although it is exposed as a NUMA node. 
As such, the CPU implements an on-chip hardware structure to facilitate fast cache coherence checks for memory accesses to the true \cxlmem.
These are important differences that may change conclusions made by prior work on the performance characteristics of \cxlmem 
and consequently the effectiveness of the proposals at the system level.
\textbf{(C3)}
The sub-NUMA clustering (SNC) mode provides LLC isolation among SNC nodes (\S\ref{sec:setup}) by directing the CPU cores within an SNC node to evict their L2 cache lines from its local memory exclusively to LLC slices within the same SNC node.  
However, when CPU cores access \cxlmem, they end up breaking the LLC isolation, as L2 cache lines from \cxlmem can be evicted to LLC slices in any SNC nodes.
%
%When CPU cores access \cxlmem, they end up breaking LLC isolation provided by the sub-NUMA clustering (SNC) mode (\S\ref{sec:setup}), in which the CPU cores in an SNC node are supposed to evict their L2 cache lines only to LLC slices within the same SNC node.  
%
%However, L2 cache lines from \cxlmem can be evicted to LLC slices in any SNC nodes.
%
Consequently, accessing \cxlmem can benefit from  effectively 2--4$\times$ larger LLC capacity than accessing local DDR memory, notably compensating for the longer latency of accessing \cxlmem for cache-friendly applications. 
%This effectively offers 2--4$\times$ larger LLC capacity for \cxlmem than local DDR memory, notably compensating for the longer latency of accessing \cxlmem for cache-friendly applications. 
%

\niparagraph{Na\"ively used \cxlmem considered harmful (\S\ref{sec:app}).}
Using a system with a \cxlmem device, we evaluate a set of applications with diverse memory access characteristics and different performance metrics (\eg, response time and throughput). 
Subsequently, we present the following \textbf{\underline{F}}indings.
\textbf{(F1)} Simple applications (\eg, key-value-store) demanding $\mu$s-scale latency are highly sensitive to memory access latency. 
Consequently, allocating pages to \cxlmem increases the tail latency of these applications by 10--82\%  compared to local DDR memory.
Besides, the state-of-the-art CXL-memory-aware page placement policy for a tiered memory system~\cite{tpp} actually increases tail latency even further when compared to statically partitioning pages between DDR memory and \cxlmem. 
This is due to the overhead of page migration.
\textbf{(F2)} Complex applications (\eg, social network microservices) exhibiting $m$s-scale latency experience a marginal increase in tail latency even when most of pages are allocated to \cxlmem. 
This is because the longer latency of accessing \cxlmem contributes marginally to the end-to-end latency of such applications.
% check
\textbf{(F3)} Even for memory-bandwidth-intensive applications, na\"ively allocating 50\% of pages to \cxlmem based on the default OS policy may result in lower throughput, despite higher aggregate bandwidth delivered by using both DDR memory and \cxlmem.

\niparagraph{CXL-memory-aware dynamic page allocation policy (\S\ref{sec:policy}).} 
To showcase the usefulness of our characterizations and findings described above, we propose \policy, a \textbf{\underline{C}}XL-memory-\textbf{\underline{a}}ware dynamic \textbf{\underline{p}}age alloca\textbf{\underline{tion}} policy for the OS to more efficiently use the bandwidth expansion capability of \cxlmem.
Specifically, \policy begins by determining the bandwidth of manufacturer-specific \cxlmem devices. % used by systems. 
Subsequently, \policy periodically monitors various CPU counters, such as memory access latency experienced by (co-running) applications and assesses the bandwidth consumed by them at runtime.
Lastly, based on the monitored CPU counter values, \policy estimates memory-subsystem performance over periods. 
When a given application demands an allocation of new pages, \policy considers the history of memory subsystem performance and the percentage of pages allocated to \cxlmem in the past. 
Then, it adjusts the percentage of the pages allocated to \cxlmem to improve the overall system throughput using a simple greedy algorithm. 
Our evaluation shows that \policy improves the throughput of a system co-running a set of memory-bandwidth-intensive SPEC CPU2017 benchmarks by 24\%, compared with the default static page allocation policy set by the OS.

\section{Background}
\label{sec:back}

\subsection{Compute eXpress Link (CXL)}
\label{sec:back:cxl}
PCIe is the industry standard for a high-speed serial interface between a CPU and I/O devices.
Each lane of the current PCIe 5.0 can deliver 32 GT/s (\eg, $\sim$64~GB/s with 16 lanes).
Built on the physical layer of PCIe, the CXL standard defines three separate protocols: \cxlio, \cxlcache, and \texttt{CXL.mem}. 
\cxlio uses protocol features of the standard PCIe, such as transaction-layer packet (TLP) and data-link-layer packet (DLLP), to initialize the interface between a CPU and a device~\cite{cxl1}. 
\cxlcache and \texttt{CXL.mem} use the aforementioned protocol features for the device to access the CPU's memory and for the CPU to access the device's memory, respectively.

The \texttt{CXL.mem} protocol accounts only for memory accesses from the CPU to the device facilitated by the Home Agent (HA) and the CXL controller on the CPU and the device, respectively~\cite{snia-cxl}. 
The HA handles the \texttt{CXL.mem} protocol and transparently exposes \cxlmem to the CPU as memory in a remote NUMA node. 
That is, the CPU can access \cxlmem with load and store instructions in the same way as it accesses memory in a remote NUMA node. 
This has an advantage over other memory expansion technologies, such as RDMA, which involves the device's DMA engine and thus has different memory access semantics. 
Lastly, %by being able to 
when the CPU accesses \cxlmem, 
%with load and store instructions, %the CPU caches 
it caches data from/to the \cxlmem in every level of its cache hierarchy. 
This has been impossible with any other memory extension technologies except for persistent memory.

\begin{figure}[!t]
    \centering
    %\vspace{-3pt}
    \includegraphics[width=\linewidth]{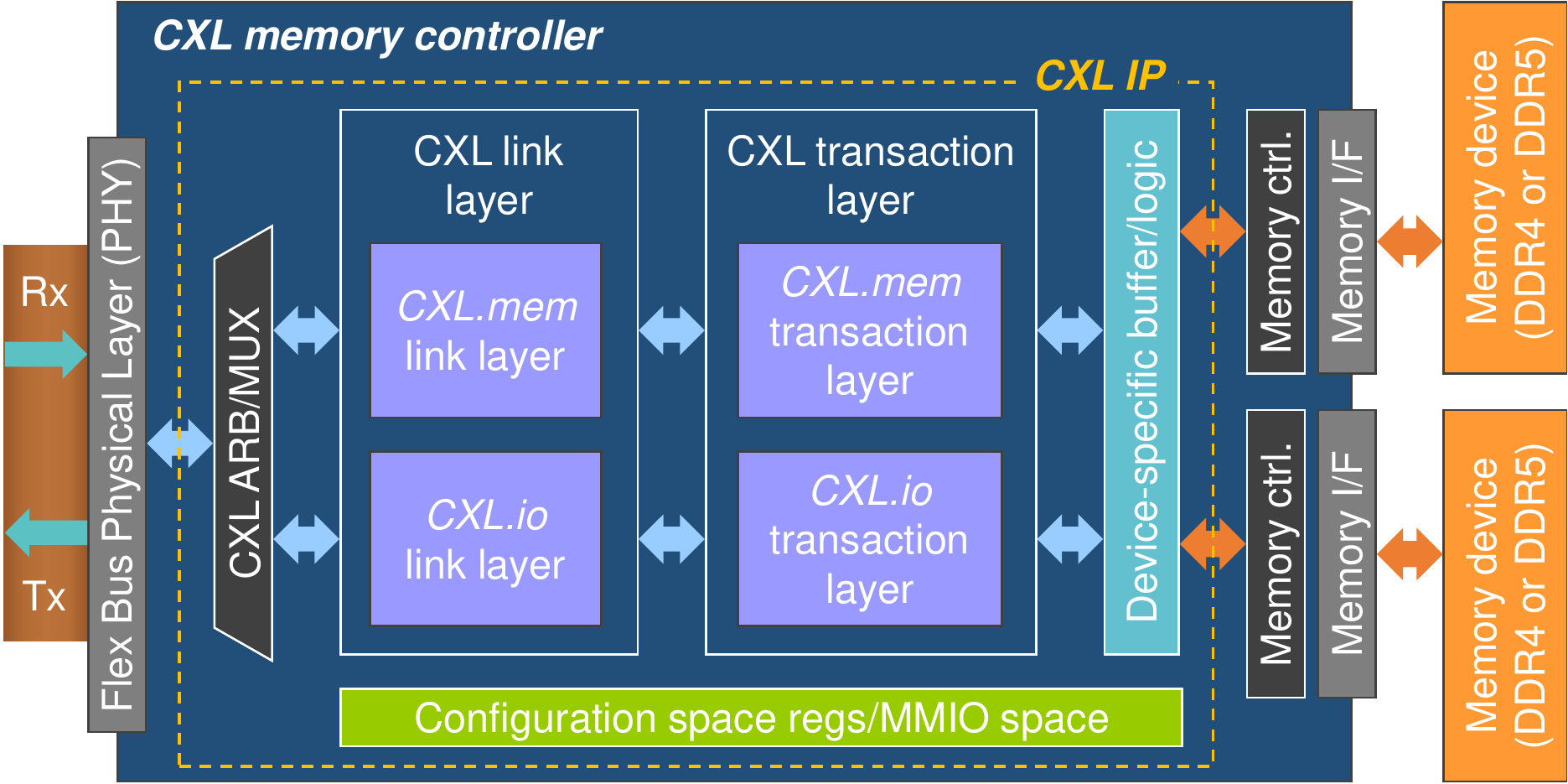}
    \vspace{-12pt}
    \caption{CXL.mem controller architecture.} 
    \label{fig:cxl-mem-ctrl}
    \vspace{-6pt}

\end{figure}

\subsection{CXL-ready Systems and Memory Devices}
\label{sec:cxl_enabled_hw}
CXL requires hardware support from both the CPU and  devices. 
Both the latest 4\textsuperscript{th}-generation Intel Xeon Scalable Processor (Sapphire Rapids) 
and the latest 4\textsuperscript{th}-generation AMD EPYC Processor (Genoa) are among the first server-class commodity CPUs to support the CXL 1.1 standard~\cite{spr, amd_4th_gen_epic}. 
\figref{fig:cxl-mem-ctrl} depicts a typical architecture of \texttt{CXL.mem} controllers.
It primarily consists of (1) PCIe physical layer, (2) CXL link layer, (3) CXL transaction layer, and (4) memory controller blocks.
(2), (3), and other CXL-related components are collectively referred to as CXL IP in this paper.
As of today, in addition to some research prototypes, multiple \cxlmem devices have been designed by major hardware manufacturers, such as Samsung~\cite{smdk_github}, SK Hynix~\cite{skhynix-cxl}, Micron~\cite{micron-cxl}, and Montage~\cite{montage-cxl}. 
To facilitate more flexible memory functionality and near-memory computing capability, Intel also enables the CXL protocol in its latest Agilex-I series FPGA~\cite{intel-agi}, integrated with hard CXL IPs to support the \cxlio, \cxlcache, and \texttt{CXL.mem}~\cite{rtile_cxl_ip}. 
Lastly, unlike a true NUMA node typically based on a large server-class CPU, a \cxlmem device does not have any CPU cores, caches, or long interconnects between the CXL IP and the memory controller in the device. 

\section{Evaluation Setup}
\label{sec:setup}
\subsection{System and Device}
\label{sec:setup:system}
\niparagraph{Systems.} We use a server to evaluate the latest commercial hardware supporting \cxlmem (Table~\ref{table:config-table}). 
The server consists of two Intel Sapphire Rapids (SPR) CPU sockets. 
One socket is populated with eight 4800~MT/s DDR5 DRAM DIMMs (128~GB) across eight memory channels. 
The other socket is populated with only one 4800~MT/s DDR5 DRAM DIMM to emulate the bandwidth and capacity of \cxlmem. %, without using the CPU cores in the socket. 
The Intel SPR CPU integrates four CPU chiplets, each with up to 15 cores and two DDR5 DRAM channels. 
A user can choose to use the 4 chiplets as a unified CPU, or each chiplet (or two chiplets) as a NUMA node in the SNC mode.
Such flexibility is to give users strong isolation of shared resources, such as LLC, among applications. 
Lastly, we turn off the hyper-threading feature and set the CPU core clock frequency to 2.1~GHz for more predictable performance.

\niparagraph{CXL memory devices.} We take three \cxlmem devices (`\cxlmem devices' in Table~\ref{table:config-table}), each featuring different CXL IPs (ASIC-based hard IP and FPGA-based soft IP) and DRAM technologies (DDR5-4800, DDR4-2400, and DDR4-3200). 
Since the CXL protocol itself does not prescribe the underlying memory technology, it can seamlessly and transparently accommodate not only DRAM but also persistent memory, flash~\cite{ms-ssd}, and other emerging memory technologies.
Consequently, various \cxlmem devices may exhibit different latency and bandwidth characteristics.

\begin{table}[t!]
\caption{System configurations.}
\label{table:config-table}
\vspace{-10pt}
\begin{center}
\resizebox{\columnwidth}{!}{%
\begin{tabular}{c||ccc}
\hline
\multicolumn{4}{c}{\textbf{Dual-socket server system}}\TBstrut\\
\hline
\textbf{Component} & \multicolumn{3}{c}{\textbf{Desription}}\TBstrut\\
\hline
OS (kernel) & \multicolumn{3}{c}{Ubuntu 22.04.2 LTS (Linux kernel v6.2)}\Tstrut\\
\multirow{2}{*}{CPU} & \multicolumn{3}{c}{2$\times$ Intel\textsuperscript{\textregistered} Xeon 6430 CPUs @2.1~GHz~\cite{6430}, 32 cores }\Tstrut\\
& \multicolumn{3}{c}{and 60~MB LLC per CPU, Hyper-Threading disabled}\\
\multirow{2}{*}{Memory} & \multicolumn{3}{l}{Socket 0: 8$\times$ DDR5-4800 channels}\Tstrut\\
& \multicolumn{3}{l}{Socket 1: 1$\times$ DDR5-4800 channel (emulated CXL memory)}\Bstrut\\
\hline
\hline
\multicolumn{4}{c}{\textbf{CXL memory devices}}\TBstrut\\
\hline
\textbf{Device} & \textbf{CXL IP} & \textbf{Memory technology} & \textbf{Max. bandwidth}\TBstrut\\
\hline
CXL-A & Hard IP & DDR5-4800 & 38.4~GB/s per channel\Tstrut\\
CXL-B & Hard IP & 2$\times$ DDR4-2400 & 19.2 GB/s per channel\\
CXL-C & Soft IP & DDR4-3200 & 25.6~GB/s per channel\Bstrut\\
\hline
\end{tabular}
}
\label{tab:smartssd}
\end{center}
\vspace{-6pt}
\end{table}

\subsection{Microbenchmark}
\label{sec:setup:microbench}
To characterize the performance of \cxlmem, we use two microbenchmarks.
First, we use Intel Memory Latency Checker (MLC)~\cite{mlc},  a tool used to measure memory latency and bandwidth for various usage scenarios.
%It also provides several options for more fine-grained investigation where b/w and latencies from a specific set of cores to caches or memory can be measured as well
%
Second, we use a microbenchmark dubbed \arch (\textbf{\underline{m}}easuring \textbf{\underline{e}}fficiency of \textbf{\underline{m}}em\textbf{\underline{o}}ry subsystems).
It shares some features with Intel MLC, but we develop it to give more control over characterizing memory subsystem performance in diverse ways.
For instance, it can measure the latency and bandwidth of a specific memory access instruction (\eg, AVX-512 non-temporal load and store instructions). 

\subsection{Benchmark}
\label{sec:setup:benchmark}
\niparagraph{Latency-sensitive applications.}
We run \texttt{Redis}~\cite{redis}, a popular high-performance in-memory key-value store, with \texttt{YCSB}~\cite{socc10:ycsb}. % as a latency-sensitive application. 
We use a uniform distribution of keys, ensuring maximum stress on the memory subsystem, unless we explicitly specify the use of other distributions. 
We also run DeathStarBench (\texttt{DSB})~\cite{deathstartbench}, an open-source benchmark suite designed to evaluate the performance of microservices. %, as another latency-sensitive application. 
It uses \texttt{Docker} to launch components of a microservice, including machine learning (ML) inference logic, web backend, load balancer, caching, and storage.
Specifically, we evaluate three \texttt{DSB} workloads: (1) \texttt{compose posts}, (2) \texttt{read user timelines}, and (3) \texttt{mixed workloads} (10\% of \texttt{compose posts}, 30\% of  \texttt{read user timelines}, and 60\% of \texttt{read home timelines}) as a social network framework. 
Lastly, we run \texttt{FIO}~\cite{GitHubax76:online}, an open-source  tool used for benchmarking storage devices and file systems, to evaluate the latency impact of using \cxlmem for OS page cache.
The page cache is supported by the standard Linux storage subsystem, which holds recently accessed storage data (\eg, files) in unused main memory space to reduce the number of accesses to slow storage devices. 

\niparagraph{Throughput applications.}
First, we run an inference application based on a deep learning recommendation model (\texttt{DLRM})  with the same setup as \texttt{MERCI}~\cite{asplos21lee}. 
The embedding reduction step in DLRM inference is known to have a large memory footprint and is responsible for 50--70\% of the inference latency~\cite{asplos21lee}. 
Second, we take the SPECrate CPU2017 benchmark suite~\cite{SPECCPU70:online}, which is commonly used to evaluate the throughput of systems in datacenters. Then we assess misses per kilo instructions (MPKI) of every benchmark and run the four benchmarks with the highest MPKI: (1) \texttt{fotonik3d}, (2) \texttt{mcf}, (3) \texttt{roms}, and (4) \texttt{cactuBSSN}.
We run multiple instances of a single benchmark or two different benchmarks.

\section{Memory Latency and Bandwidth Characteristics}
\label{sec:character}
In this section, we first evaluate the latency and bandwidth of accessing different memory devices: an emulated \cxlmem device based on DDR5 memory in a remote NUMA node (DDR5-R), and three true \cxlmem devices (CXL-A, CXL-B, and CXL-C). 
We conduct this evaluation to understand the performance characteristics of various \cxlmem devices for different memory access instruction types. %, and then to understand key differences among them. 
Next, we investigate interactions between the Intel SPR CPU's cache hierarchy and the \cxlmem devices.
%
%We conduct these evaluations and explorations to understand the performance characteristics of various \cxlmem devices for different memory access instruction types. %, and then to understand key differences among them. 
%

\subsection{Latency}
\label{sec:mem-charac:latency}
\begin{figure}[b]
    \centering
        \vspace{-6pt}

\includegraphics[width=\linewidth]{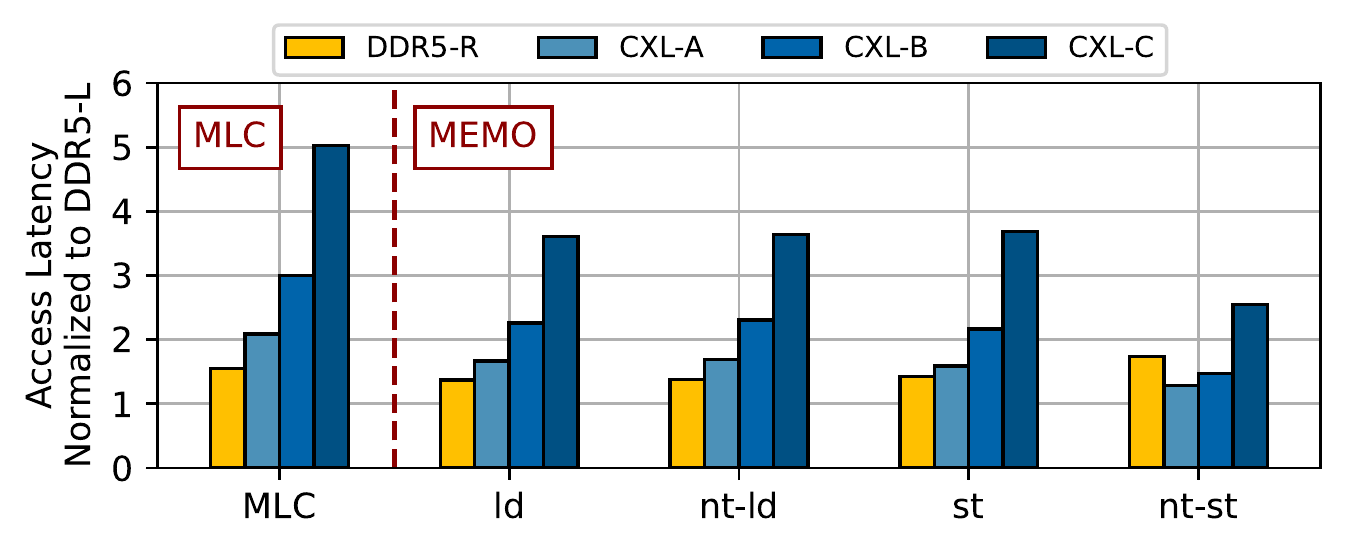}
    \vspace{-18pt}
    \caption{Random memory access latency of various memory devices (DDR5-R, CXL-A, CXL-B, and CXL-C), measured by Intel MLC and \arch. 
    %All the latency values are normalized to those of local DDR5 memory (DDR5-L). 
    %Both DDR5-L and DDR5-R use only a single DRAM channel to emulate the bandwidth of a \cxlmem device. 
    %\texttt{st} instructions are followed by a cache line write-back instruction (\texttt{clwb}) to ensure that the stored value is written all the way back to memory. \texttt{nt-st} instructions are followed by \texttt{mfence} to confirm the completion of \texttt{nt-st}. The presented store latency is measured and perceived by software, and the true store latency  can be longer.
    } 
    \label{fig:seq_latency}
\end{figure}

\figref{fig:seq_latency} presents the measured latency values of accessing both emulated and true \cxlmem devices. % across four memory access instruction types.
The first group of bars shows average (unloaded idle) memory access latency values measured by Intel MLC that performs pointer-chasing (\ie, getting the memory address of a load from the value of the preceding load) in a memory region larger than the total LLC capacity of the CPU. 
This effectively measures the latency of serialized memory accesses.
The remaining four groups of bars show the average memory access latency values measured by \arch for four memory access instruction types: (1) temporal load (\texttt{ld}), (2) non-temporal load (\texttt{nt-ld}), (3) temporal store (\texttt{st}), and (4) non-temporal store (\texttt{nt-st}). 
For these groups, we first execute \texttt{clflush}  to flush all cache lines from the cache hierarchy and then \texttt{mfence} to ensure the completion of flushing the cache lines.
Then, we execute 16 memory access instructions back to back to 16 random memory addresses in a cacheable memory region.
To measure the execution time of these 16 memory access instructions, we execute \texttt{rdtsc}, which reads the current value of the CPU’s 64-bit time-stamp counter into a register immediately before and after executing the 16 memory access instructions, followed by an appropriate fence instruction.
This effectively measures the latency of random parallel memory accesses for each memory access instruction type for a given memory device.
To obtain a representative latency value, we repeat the measurement 10,000 times and choose the median value to exclude outliers caused by TLB misses and OS activities.  

Analyzing the latency values shown in \figref{fig:seq_latency}, we make the following \textbf{\underline{O}}bservations. 

\niparagraph{(O1) The full-duplex CXL and UPI interfaces reduce memory access latency.} 
\arch gives emulated \cxlmem 76\% lower \texttt{ld} latency than Intel MLC. 
This difference arises because serialized memory accesses by Intel MLC cannot exploit the full-duplex capability of the UPI interface that connects NUMA nodes.
In contrast, random parallel memory accesses by \arch can send memory commands/addresses and receive data in parallel through the full-duplex UPI interface, effectively halving the average latency cost of going through the UPI interface.
Since true \cxlmem is also based on the full-duplex interface (\ie, PCIe), it enjoys the same benefit as emulated \cxlmem. 
%
%Nonetheless, CXL-A gives 3 percentage points lower \texttt{ld} latency than DDR5-R with \arch. 
Nonetheless, with \arch, CXL-A gets a 3 percentage points more \texttt{ld} latency reduction than for DDR5-R. 
(O3) explains the reason for this additional latency reduction.  

\niparagraph{(O2) The latency of accessing true \cxlmem devices is highly dependent on a given CXL controller design.} 
\figref{fig:seq_latency} shows that CXL-A exhibits only 35\% longer \texttt{ld} latency than DDR5-R, while CXL-B and CXL-C present almost 2$\times$ and 3$\times$ longer \texttt{ld} latency, respectively.
Even with the same DDR4 DRAM technology, CXL-C based on DDR4-3200 gives 67\% longer \texttt{ld} latency than CXL-B based on DDR4-2400.

\niparagraph{(O3) Emulated \cxlmem can give longer memory access latency than true \cxlmem.} 
When issuing memory requests to emulated \cxlmem, the local CPU  must first check with the remote CPU, which is connected through the inter-chip UPI interface, for cache coherence~\cite{loughlin2022moesi,molka2015cache}. 
Moreover, the memory requests must travel through a long intra-chip interconnect within the remote CPU to reach its memory controllers~\cite{icpe22-velten}. 
% check
%Moreover, the memory requests must travel through a long interconnect path between the UPI interface and memory controllers in the remote CPU~\cite{icpe22-velten}. 
% check
These overheads increase with more CPU cores, \ie, more caches and a longer interconnect path. % check 
For example, %when measuring 
the \texttt{ld} latency values of DDR5-R with 26- and 40-core Intel SPR CPUs are 29\% lower and 19\%  higher, respectively, than those of DDR5-L with the 32-core Intel SPR CPU used for our primary evaluations. %(\cf Table~\ref{table:config-table}).  
% check
In contrast, true \cxlmem has neither caches nor CPU cores that modify caches, although it is exposed as a remote NUMA node.
% check
%That is, the local CPU has no remote CPU to check with for cache coherence for the memory address space provided by true \cxlmem.
%
As such, the CPU %supporting true \cxlmem 
implements an on-chip hardware structure to facilitate fast cache coherence checks for memory accesses to the true \cxlmem.
% check
Moreover, true \cxlmem features a short intra-chip interconnect within the CXL controller to reach its memory controllers.
%path between the CXL IP and the memory controller in the CXL controller. 
% check

Specifically for DDR5-R and CXL-A, \arch provides 76\% and 79\% lower \texttt{ld} latency values, respectively, than Intel MLC. 
% check
Although both DDR5-R and CXL-A benefit from (O1), CXL-A gives a more \texttt{ld} latency reduction than DDR5-R. 
% check
This arises from the following differences between \arch and Intel MLC. 
% check
As Intel MLC accesses memory serially, the local CPU performs the aforementioned cache coherence checks one by one. 
% check
By contrast, \arch accesses memory in parallel. 
% check
In such a case, memory accesses to emulated \cxlmem incur a burst of cache coherence checks that need to go through the inter-chip UPI interface, %and increase the pressure on the snoop filter in the remote socket. %, 
leading to cache coherence traffic congestion. 
% check
This, in turn, increases the time required for cache coherence checks. 
% check
%a burst of cache coherence checks of which the response time is usually larger than the latency of accessing DRAM~\cite{mccalpin2018topology}. 
%
However, memory accesses to true \cxlmem suffer notably less from this overhead because the  CPU checks cache coherence through its local on-chip structure described earlier. 
% check
This contributes to a further reduction in the \texttt{ld} latency for true \cxlmem. %between emulated \cxlmem and true \cxlmem. % check
%
%Since memory accesses to true \cxlmem do not suffer from such overhead, the \texttt{ld} latency gap between emulated \cxlmem and true \cxlmem is further reduced. 
%
%These explanations are supported by two other experiments. % check
%
%First, when the number of parallel memory accesses is doubled, % (\ie, 16 to 32), 
%the \texttt{ld} latency gap between \arch and Intel MLC  increases from 3 percentage points to 2 percentage points. %check 
%
%Second, DDR5-R, based on a 40-core Intel SPR CPU, presents 4.2\% longer \texttt{ld} latency than CXL-A due to a higher overhead of cache coherence checks. % check
%
%These explanations are supported by another experiment.
%
Also note that DDR5-R, based on a 40-core Intel SPR CPU, presents 4\% longer \texttt{ld} latency than CXL-A due to a higher overhead of cache coherence checks. % check

The overhead of cache coherence checks becomes even more prominent for \texttt{st} because of two reasons.
% check
First, the \texttt{st} latency is much higher than the \texttt{ld} latency in general.
% check
For example, the latency of \texttt{st} to DDR5-R is 2.3$\times$ longer than that of \texttt{ld} from DDR5-L.
% check
This is because of the cache write-allocate policy in Intel CPUs. 
% check
When \texttt{st} experiences an LLC miss, the CPU first reads 64-byte data from memory to a cache line (\ie, implicitly executing \texttt{ld}), and then writes modified data to the cache line~\cite{perf-analysis-guide-i7}.
% check
This overhead is increased for both emulated \cxlmem and true \cxlmem, as the overhead of traversing the UPI and CXL interfaces is doubled.
% check
Yet, the latency of \texttt{st} to emulated \cxlmem increases more than that of \texttt{st} to emulated \cxlmem, when compared to \texttt{ld} or \texttt{nt-ld}. 
% check
This is because the emulated \cxlmem incurs a higher overhead for cache coherence checks than the true \cxlmem, as discussed earlier.
% check.

Lastly, although both \texttt{nt-ld} and \texttt{nt-st} bypass caches and directly access memory, the local CPU accessing emulated \cxlmem still needs to check with the remote CPU for cache coherence~\cite{Inter64}. 
% check
This explains why the \texttt{nt-ld} latency values of all the memory devices are similar to those of \texttt{ld} that needs to be served by memory in \figref{fig:seq_latency}.
% check
Unlike \texttt{st}, however, \texttt{nt-st} does not read 64-byte data from the memory since it does not allocate a cache line by its semantics, which eliminates the memory access and cache coherence overheads associated with implicit \texttt{ld}. 
% check
Therefore, the absolute values of \texttt{nt-st} latency across all the memory devices are smaller than those of \texttt{st} latency. 
% check
Furthermore, \texttt{nt-st} can also offer shorter latency than \texttt{ld} and \texttt{nt-ld} because the CPU issuing \texttt{nt-st} sends the address and data simultaneously.
In contrast, the CPU issuing \texttt{ld} or \texttt{nt-ld} sends the address first and then receive the data later, which makes signals go through the UPI and CXL interfaces twice. 
With shorter latency for a cache coherency check, \texttt{nt-st} to true \cxlmem can be shorter than \texttt{nt-st} to emulated \cxlmem.
For instance, CXL-A exhibits a 25\% lower latency than DDR5-R for \texttt{nt-st}.
Note that \texttt{nt-st} behaves differently depending on whether the allocated memory region is cacheable or not~\cite{Inter64}, and we conduct our experiment with a cacheable memory region.
% check
%This explains why the \texttt{nt-st} latency relative to DDR5-L is lower than the \texttt{st} latency.
% check

\subsection{Bandwidth}
\label{sec:mem-charac:bandwidth}
\begin{figure}[t]
    %\vspace{-6pt}
    \centering
    \begin{subfigure}[b]{\linewidth}
        \centering
        \includegraphics[width=\linewidth]{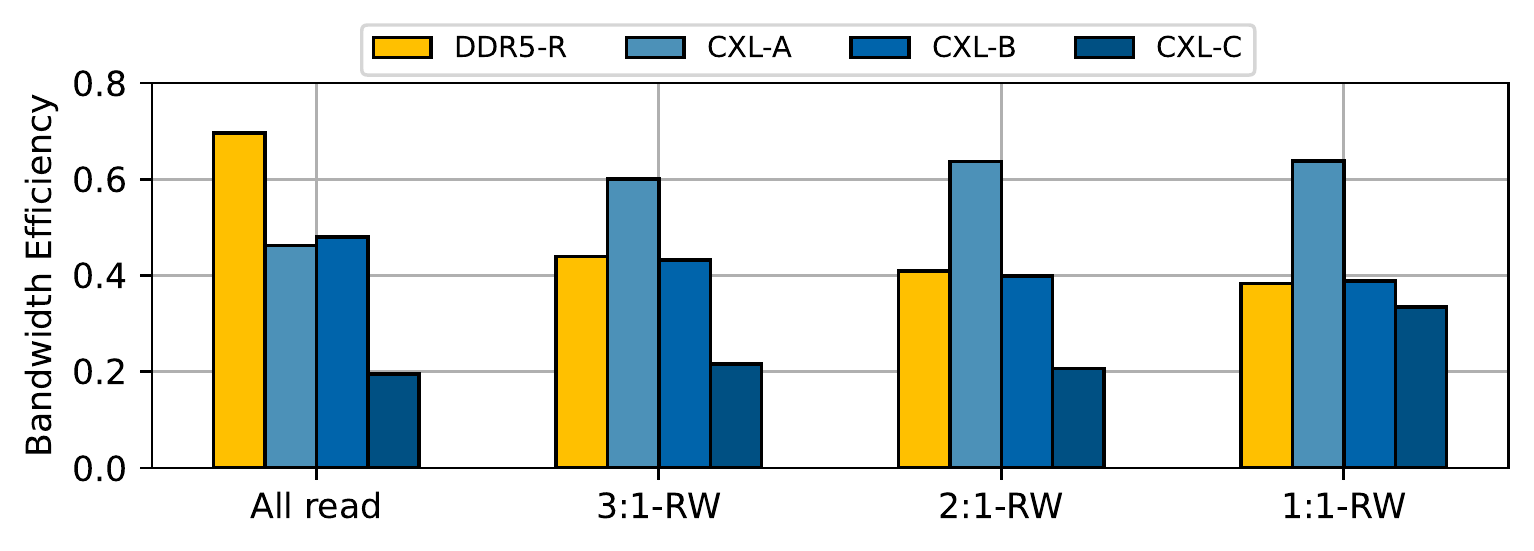}
        \vspace{-12pt}
        \caption{MLC with various read and write ratios
        %\yan{All read, 3:1, 2:1, 1:1, stream-traid. Bars, group by devices, 1st norm to max mem b/w, 2nd norm to DDR-L1}
        }
        \label{subfig:bw_eff_mlc}
    \end{subfigure}
    \vfill
    \begin{subfigure}[b]{\linewidth}
        \centering
        \includegraphics[width=\linewidth]{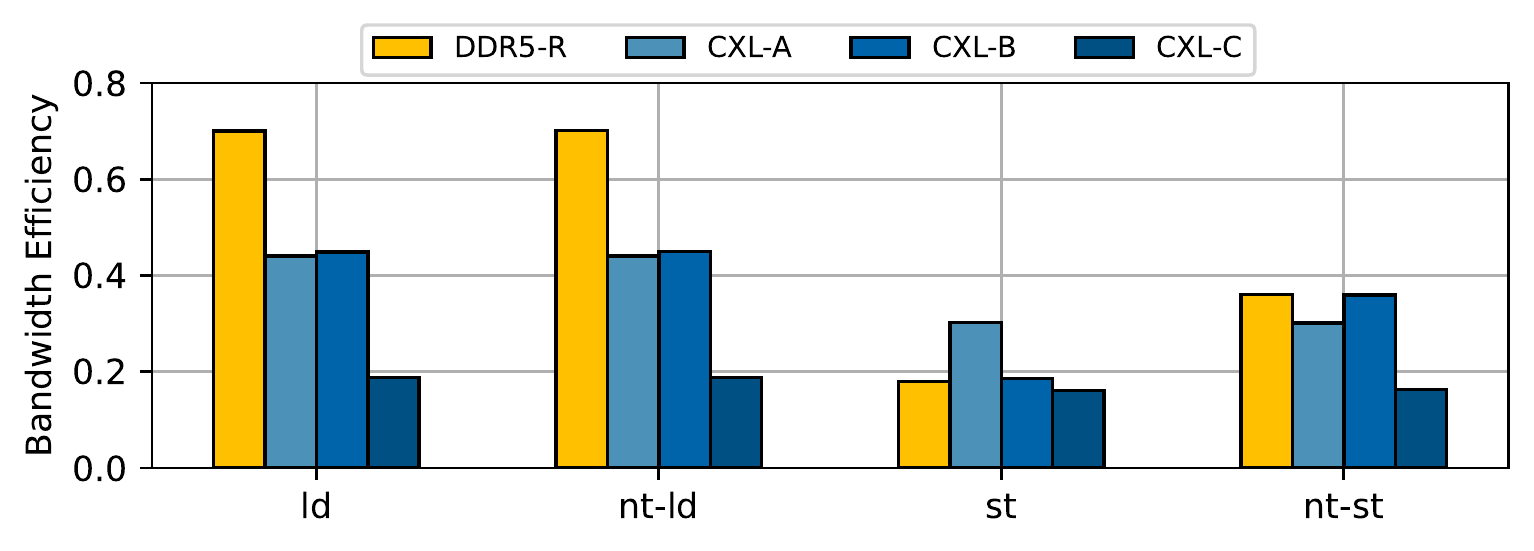}
        \caption{\arch with different memory instruction types}
        \label{subfig:bw_eff_memo}
    \end{subfigure}
    \vspace{-15pt}
    \caption{Efficiency of maximum sequential memory access bandwidth across different memory types.}
    \label{fig:seq_bandwidth}
    \vspace{-6pt}
\end{figure}

The sequential memory access bandwidth represents the maximum throughput of the memory subsystem when all the CPU cores sends memory requests in this paper.
% check
Nonetheless, it notably varies across (1) CXL controller designs, (2) memory access instruction types, and (3) DRAM technologies (\ie, DDR5-4800, DDR4-3200, and DDR4-2400 in this paper).
% check
As such, for fair and insightful comparison, we use bandwidth efficiency as a metric, normalizing the measured bandwidth to the theoretical maximum bandwidth. 
%of a given memory device for each memory access instruction type, 
%(\eg, `All Read' in \figref{subfig:bw_eff_mlc} and \texttt{ld} in \figref{subfig:bw_eff_memo}), 
% using the following equation:
% check
%\begin{align}
%\frac{Max\_Bandwidth_{Measured}(Memory\_Type,\;Case)}{Max\_Bandwidth_{Theoretical}(Memory\_Type)}
%\label{eq:1}
%\end{align}
%
%\figref{subfig:bw_eff_mlc} and \figref{subfig:bw_eff_memo} present 
\figref{fig:seq_bandwidth} presents the bandwidth efficiency values of DDR5-R, CXL-A, CXL-B, and CXL-C for various read/write ratios and different memory instruction types, respectively.
% check
Analyzing these values, %the bandwidth efficiency values from \figref{fig:seq_bandwidth},
we make the following \textbf{\underline{O}}bservations. 
% check

\niparagraph{(O4) The bandwidth is strongly dependent on the efficiency of CXL controllers.}
The maximum sequential bandwidth values that can be provided by the DDR5-4800, DDR4-3200, and DDR4-2400 DRAM technologies are 38.4~GB/s, 25.6~GB/s, and 19.2~GB/s per channel, respectively.
% check
%Nonetheless, both Intel MLC `All Read' and \arch \texttt{ld} show that 
Nonetheless, \figref{subfig:bw_eff_mlc} shows that DDR5-R, CXL-A, CXL-B, and CXL-C provides only 70\%, 46\%, 47\%, and 20\% of the theoretical maximum bandwidth, respectively, for `All Read'.
% check
%The bandwidth efficiency gap between DDR5-L and DDR5-R based on the same memory controller is \textcolor{red}{X} percentage points, representing the overhead of accessing remote memory.
%
DDR5-R and CXL-A are based on the same DDR5-4800 DRAM technology, yet the bandwidth efficiency of DDR5-R is 23 percentage points higher than that of CXL-A.
% check
We speculate that the lower efficiency of the CXL-A's memory controller for memory read accesses contributes to this bandwidth efficiency gap, as both DDR5-R and CXL-A exhibit similar \texttt{ld} latency values.
% check

As the write ratio increases, however, CXL-A starts to provide higher bandwidth efficiencies.
% check
For example, \figref{subfig:bw_eff_mlc} shows that the bandwidth efficiency of CXL-A for `2:1-RW' is 23 percentage points higher than that of DDR5-R.
% check
We speculate that the CXL-A's memory controller is designed to handle interleaved memory read and write accesses more efficiently than the DDR5-R's and CXL-B's memory controllers.  
% check
This is supported by (1) the fact that \texttt{st} involves both memory read and write accesses due to implicit \texttt{ld} when it incurs a cache miss (\cf (O3)), and (2) 
%the bandwidth efficiency of CXL-A for \texttt{st} is 12.2 and \texttt{red}{X} percentage points 
%higher than that of DDR5-R and CXL-B, whereas 
the bandwidth efficiency of CXL-A for all the other memory access instruction types is lower than that of DDR5-R and CXL-B.
% check
This also implies that the higher bandwidth efficiency of CXL-A for \texttt{st} is not solely attributed to a unique property of true \cxlmem. %, as it is 11.6 percentage points higher than that of CXL-B that . 
% check

\figref{subfig:bw_eff_memo} shows that the bandwidth efficiency of CXL-B is higher than that of CXL-A, except for \texttt{st}, although the latency values of CXL-B is higher than those of CXL-A.
% check
Specifically, the bandwidth efficiency values of CXL-B for \texttt{ld}, \texttt{nt-ld}, and \texttt{nt-st} are 1, 1, and 6 percentage points higher than that of CXL-A, respectively.
% check
We speculate that the recently developed third-party DDR5 memory controllers may not be as efficient as the mature and highly optimized DDR4 memory controller used in CXL-B for read- or write-only memory accesses.
% check
Note that CXL-C is based on DDR4-3200 DRAM technology, but it generally exhibits poor bandwidth efficiency due to the FPGA-based implementation of the CXL controller.
% check
The bandwidth efficiency values of CXL-C for \texttt{ld}, \texttt{nt-ld}, \texttt{st}, and \texttt{nt-st} are 26, 26, 3, and 20 percentage points lower, respectively, than those of CXL-B, which is based on the same DDR4 DRAM technology but provides 25\% lower theoretical maximum bandwidth per channel.
% check

\niparagraph{(O5) True \cxlmem can offer competitive bandwidth efficiency for the store compared to emulated \cxlmem.}
% check
\figref{subfig:bw_eff_memo} shows that 
\texttt{st} yields %up to 74.3\% 
lower bandwidth efficiency values than \texttt{ld} across all the memory devices due to the overheads of  implicit \texttt{ld} and cache coherence checks (\cf (O3)).
% check
Specifically, \texttt{st} to DDR5-R, CXL-A, CXL-B, and CXL-C offers 74\%, 31\%, 59\%, and 15\% lower bandwidth efficiency values, respectively, than  \texttt{ld} from DDR5-R, CXL-A, CXL-B, and CXL-C.
% check
This suggests that emulated \cxlmem experiences a notably more bandwidth efficiency degradation than true \cxlmem partly because it suffers more from the overhead of cache coherence checks.
% check
As a result, the bandwidth efficiency values of CXL-A and CXL-B for \texttt{st} are 12 and 1 percentage points higher, respectively, than DDR5-R.
% check
For \texttt{nt-st}, the bandwidth efficiency gap between emulated \cxlmem and true \cxlmem is noticeably reduced compared to \texttt{nt-ld}.
% check
Specifically, the bandwidth efficiency gap between DDR5-R and CXL-A for \texttt{nt-ld} is 26 percentage points, whereas it is reduced to 6 percentage points for \texttt{nt-st}.
% check
CXL-B provides almost the same bandwidth efficiency as DDR5-R for \texttt{nt-st}.

% check
%Although CXL-A presents an exceptionally higher bandwidth efficiency than all other memory devices, Combined with the reasons discussed in \textbf{(O4)}, CXL-B gives 22\% and 17\% higher bandwidth efficiency for \texttt{ld} and \texttt{st}, respectively, than DDR5-R, although DDR5-R still provides higher absolute bandwidth than CXL-B. 

%Note that, we also measure the random memory access bandwidth of these memory types and observe a similar trend. The only difference is \texttt{nt-st}, where for all \cxlmem devices, the bandwidth drops after using more threads to conduct the test. This is because of the limited capacity of the CPU's HA, where coherence information is monitored and tracked~\cite{loughlin2022moesi}. As \texttt{nt-st} does not occupy the core resources, more \texttt{nt-st} instructions can be issued concurrently than the regular \texttt{st}. When the number of on-the-fly \texttt{nt-st} instructions, containing random address information, exceeds the HA capacity, they will be serialized to make sure they can be correctly parsed and routed. This degrades aggregate bandwidth consumption.

\subsection{Interaction with Cache Hierarchy}
\label{sec:mem-charac:cache}

\begin{figure}[b!]
    \centering
    %\vspace{-3pt}
    \includegraphics[width=0.88\linewidth]{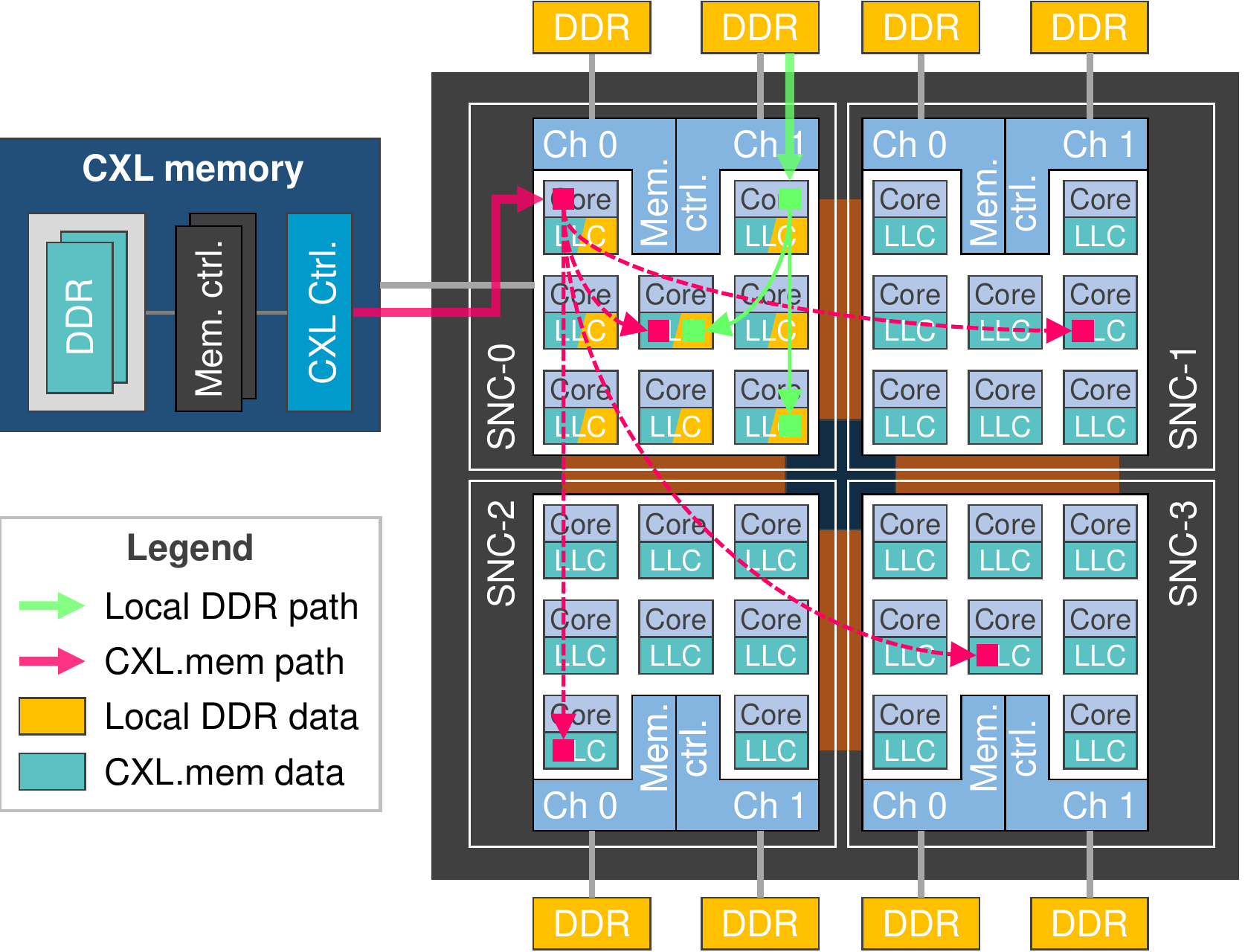}
    \caption{Difference in L2 cache line eviction paths between local DDR memory and \cxlmem in SNC mode. 
    %L2 cache lines from DDR memory are evicted only to LLC slices (light-green arrows) in the same SNC node while ones from \cxlmem are evicted to the LLC slices on one of the SNC nodes (red-dashed arrows).
    } 
    %\vspace{-12pt}
    \label{fig:spr_interation_cache_hierarchy}
\end{figure}

Starting with the Intel Skylake CPU, Intel has adopted non-inclusive cache architecture~\cite{non-inclusive-llc}.
% check
Suppose that a CPU core with non-inclusive cache architecture incurs an LLC miss that needs to be served by memory.
% check
Then it loads data from the memory into a cache line in the CPU core's (private) L2 cache rather than the (shared) LLC, in contrast to a CPU core with inclusive cache architecture.
% check
When evicting the cache line in the L2 cache, the CPU core will place it into the LLC. That is, the LLC serves as a victim cache.
The SNC mode (\S\ref{sec:setup:system}), however, restricts where the CPU core places evicted L2 cache lines within the LLC to provide LLC isolation among SNC nodes. 
% check
The LLC comprises as many slices as the number of CPU cores, %each coupled with an L2 cache, 
and
% check
%That is, 
L2 cache lines in an SNC node are evicted only to LLC slices within the same SNC node when data in the cache lines are from the local DDR memory of that SNC node (light-green lines in \figref{fig:spr_interation_cache_hierarchy}).
% check
%This provides LLC isolation among SNC nodes.
% check
In contrast, we notice that L2 cache lines can be evicted to any LLC slices within any SNC nodes when the data are from remote memory, including both emulated \cxlmem and true \cxlmem (red-dashed lines in \figref{fig:spr_interation_cache_hierarchy}). 
% check
As such, CPU cores accessing \cxlmem break LLC isolation among SNC nodes in SNC mode. 
%
%Nonetheless, they 
This makes such CPU cores benefit from 2--4$\times$ larger LLC capacity than the ones accessing local  DDR memory, notably compensating for the slower access latency of \cxlmem.
% check

To verify this, we run a single instance of Intel MLC on an idle CPU and measure the average latency of randomly accessing  32~MB buffers allocated to DDR5-L and CXL-A, respectively, in the SNC mode. 
% check
The total LLC capacity of the CPU with four SNC nodes in our system is $\sim$60~MB.
% check
A 32~MB buffer is larger than the total LLC capacity of a single SNC node but smaller than that of all four SNC nodes.
% check
This shows that accessing the buffer allocated to CXL-A gives an average memory access latency of 41~ns, whereas accessing the buffer allocated to DDR5-L offers an average memory access latency of 76.8~ns. 
% check
The shorter latency of accessing the buffer allocated to CXL-A  evidently shows that the CPU cores accessing \cxlmem can benefit from larger effective LLC capacity than the CPU cores accessing local DDR memory in the SNC mode.

\niparagraph{(O6) \cxlmem interacts with the CPU's cache hierarchy differently compared to local DDR memory.} 
As discussed above, the CPU cores accessing \cxlmem are exposed to a larger effective LLC capacity in the SNC mode.
This often significantly impacts LLC hit/miss and interference characteristics that the CPU cores experience, and thus affecting the performance of applications (\S\ref{sec:app:cache}).
Therefore, we must consider this attribute of \cxlmem when analyzing the performance of applications using \cxlmem, especially in the SNC mode.
\begin{figure*}[t]
    \includegraphics[width=\textwidth]{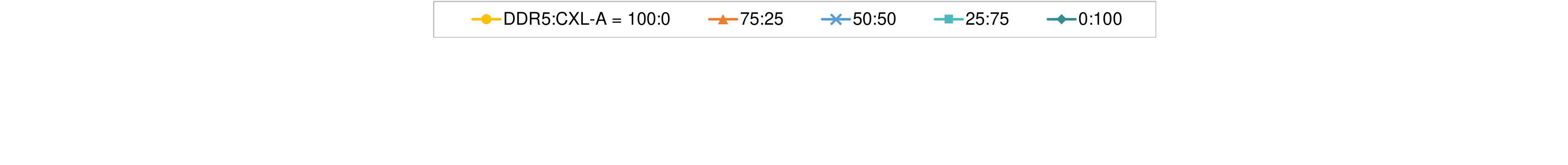}
\end{figure*}
\begin{figure*}[t]
    \vspace{-46pt}
     \begin{subfigure}[b]{0.24\textwidth}
         \centering
         \includegraphics[width=\linewidth]{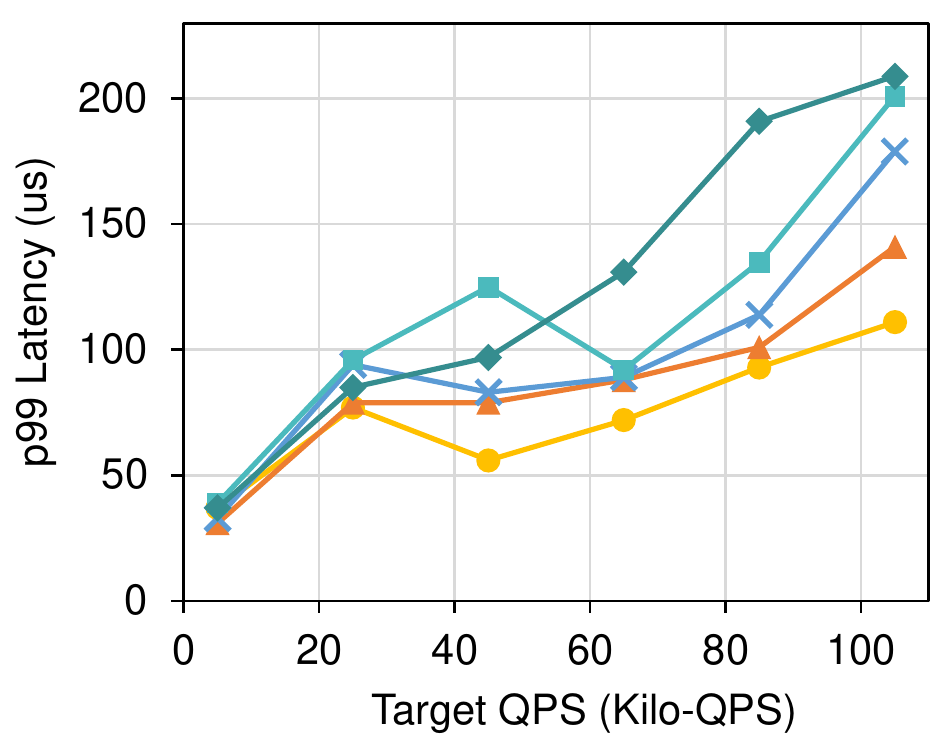}
         \caption{[Redis] YCSB-A (RD:UPD = 5:5)}
         \label{fig:ycsb_redis}
     \end{subfigure}
     \hfill
     \begin{subfigure}[b]{0.24\textwidth}
         \centering
         \includegraphics[width=\linewidth]{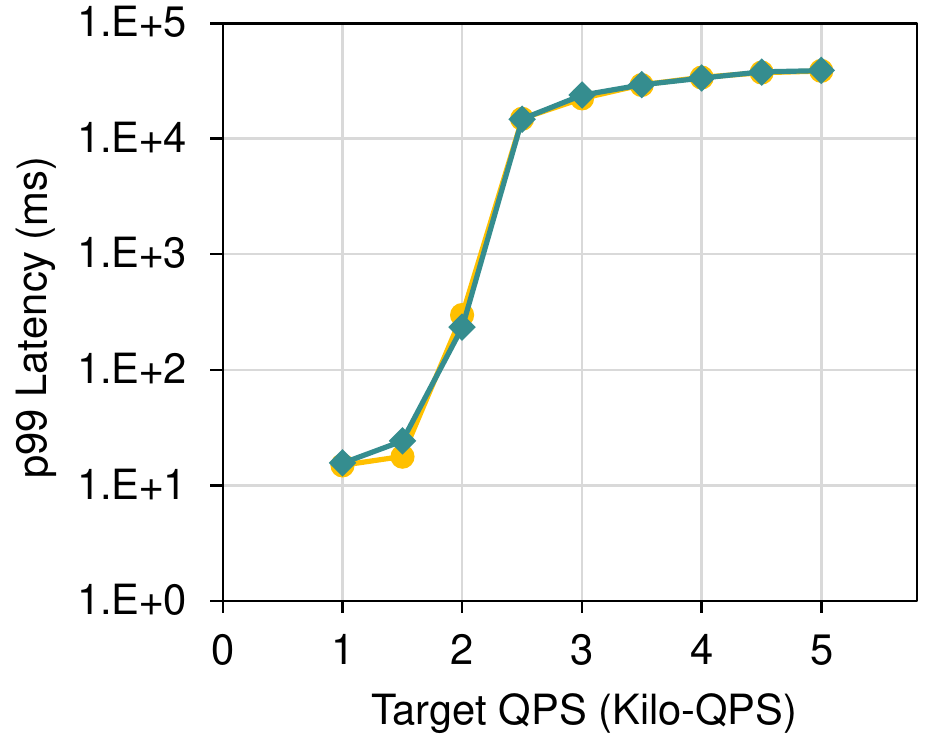}
         \caption{[DSB] compose posts}
         \label{fig:dsb_compose}
     \end{subfigure}
     \hfill
     \begin{subfigure}[b]{0.24\textwidth}
         \centering
         \includegraphics[width=\linewidth]{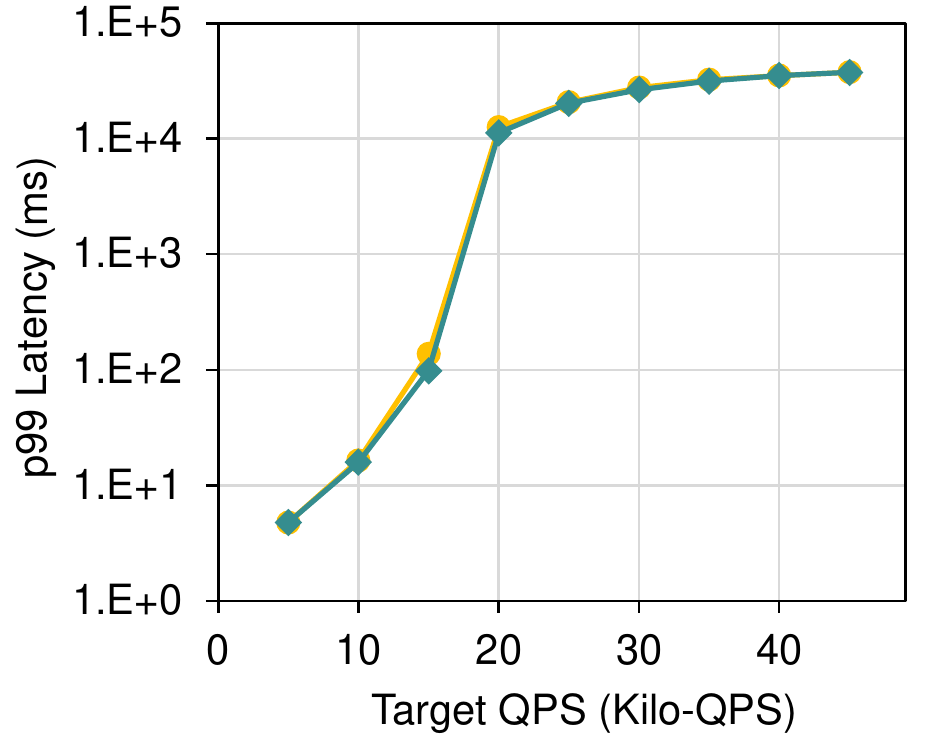}
         \caption{[DSB] read user timelines}
         \label{fig:dsb_read_user}
     \end{subfigure}
     \hfill
     \begin{subfigure}[b]{0.24\textwidth}
         \centering
         \includegraphics[width=\linewidth]{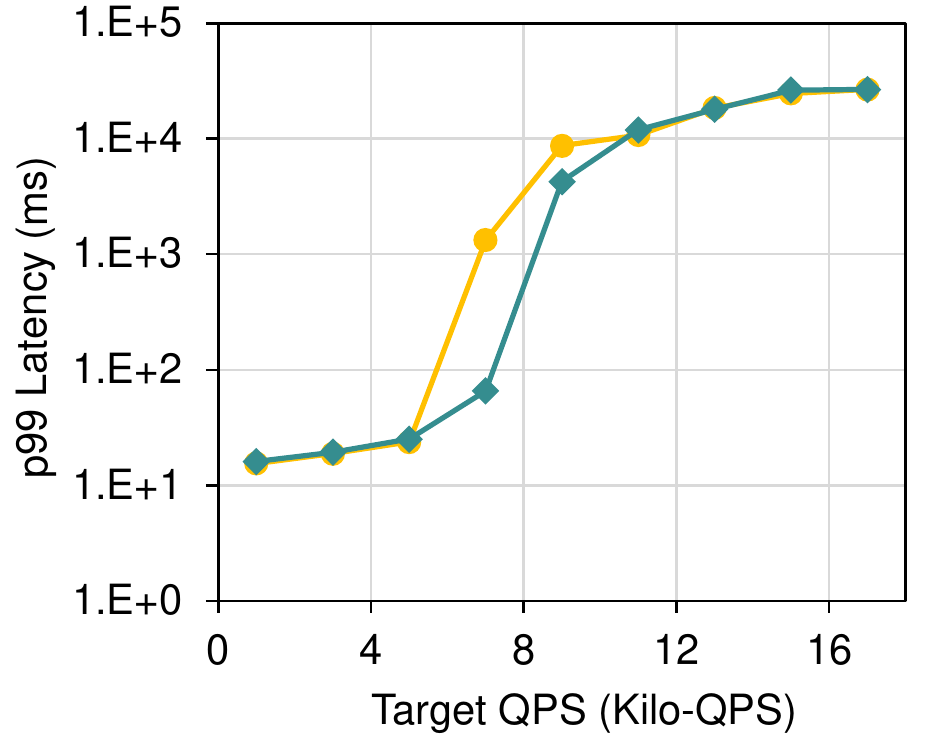}
         \caption{[DSB] mixed workloads}
         \label{fig:dsb_mixed}
     \end{subfigure}
    \vspace{-6pt}
    \caption{99\textsuperscript{th}-percentile (p99) latency values of \texttt{Redis} with various percentages of pages allocated to \cxlmem and \texttt{DSB} with 100\% of only `caching and storage' pages allocated to either \cxlmem or DDR memory.  
    %varying page interleaving ratios between DDR and \cxlmem devices.%\yan{All figures are up-to-date.} \yan{p99 gap is gone for 6b, TODO, need to re-explain} \yan{6D seems to show CXL offers some benefit (offloading DRAM pressure) -- re-tested, seems to show the same pattern} %; each bar in \figref{fig:ycsb_redis} plots the difference between targeted and delivered QPS values.
    %For this experiment, we enable the SNC mode to use two local DDR5 channels and one CXL channel to balance the bandwidth ratio between DDR5 and CXL.
    }
    %, following a guideline from Intel 
     
    %The lines in \figref{fig:ycsb_redis} are p99 latency values, and each bar is the difference between targeted and delivered requests.}
    %} 
    \label{fig:app:latency}
    %\vspace{-6pt}
\end{figure*}

\section{Impact of Using CXL Memory on Application Performance}
\label{sec:app}

To study the impact of using \cxlmem on the performance of applications (\S\ref{sec:setup:benchmark}), we take CXL-A, which provides the most balanced latency and bandwidth characteristics among the three \cxlmem devices.
% check
We use \texttt{numactl} in Linux to allocate memory pages of a given program, either fully or partially, to \cxlmem, 
%
%\cxlmem by binding the memory space demanded by a given application fully or partially with the \cxlmem space. 
% check
%For such allocation of pages to \cxlmem, 
%we use \texttt{numactl} in Linux, 
exploiting the fact that the OS recognizes \cxlmem as memory in a remote NUMA node by the OS.
% check
Specifically, \texttt{numactl} allows users to: (1) bind a program to a specific memory node (\texttt{membind} mode); (2) allocate memory pages to the local NUMA node first, and then other remote NUMA nodes only when the local NUMA node runs out of memory space (\texttt{preferred} mode); or (3) allocate memory pages evenly across a set of nodes (\texttt{interleaved} mode). %, \ie, allocating 50\% of memory pages to CXL and DDR5-L, respectively.
% check
A recent Linux kernel patch~\cite{mn_il_patch} enhances the \texttt{interleaved} mode to facilitate fine-grained control over the allocation of a specific percentage of pages to a chosen NUMA node. % (\eg, local DDR memory and \cxlmem). 
% check
For example, we can change the percentage of pages allocated to \cxlmem from the default 50\% to 25\%.
%of pages to \cxlmem, 
That is, 75\% of pages are allocated to local DDR5 memory. %adjusting the page allocation ratio between DDR and CXL from 1:1 to 3:1. 
% check

In this section, we will vary 
%the page allocation ratios between DDR memory and 
the percentage of pages allocated to
\cxlmem and analyze 
%the impact of these ratios 
its impact on performance using application-specific performance
%metrics. 
metrics,
% check
%This will set 
setting
the stage for our \cxlmem-aware dynamic page allocation policy (\S\ref{sec:policy}).
% check
Note that we enable the SNC mode to use only two local DDR5 memory channels along with one \cxlmem channel. 
% check
This is because our system can accommodate only one \cxlmem device, and it needs to make a meaningful contribution to the total bandwidth of the system. 
% check
In such a setup, the local DDR5 memory with two channels 
provides $\sim$2$\times$ higher bandwidth for \texttt{st} and $\sim$3.4$\times$ higher bandwidth for \texttt{ld} than \cxlmem. 
% check
%Still, our setup is aligned with the optimal DDR:CXL ratio suggested by multiple hardware manufacturers. %~\cite{intel-hetero-cxl, 10.1145/3533737.3535090}.
% check
As future platforms accommodate more \cxlmem devices, we may connect up to four \cxlmem devices to a single CPU socket with eight DDR5 memory channels, providing the same DDR5 to CXL channel ratio as our setup.
% check

\subsection{Latency}
\label{sec:app:latency}

\niparagraph{Redis.}
\figref{fig:ycsb_redis} shows the 99$^{th}$-percentile (p99) latency  values of \texttt{Redis} with \texttt{YCSB} workload \texttt{A} (50\% read and 50\% update) while varying the percentage of pages allocated to 
%page allocation ratio between 
\cxlmem or local DDR5 memory (referred to as DDR memory hereafter).
% check
First, allocating 100\% of pages to \cxlmem (CXL 100\%) significantly increases the p99 latency compared to allocating 100\% of pages to DDR memory (DDR 100\%), especially at high target QPS values.
% check
For example, CXL 100\% results in 10\%, 73\%, and 105\% higher p99 latency than DDR 100\% at 25~K, 45~K, and 85~K target QPS, respectively.
% check
Second, as more pages are allocated to \cxlmem, the p99 latency increases proportionally.
For instance, at 85~K target QPS, allocating 25\%, 50\%, and 75\% of pages to \cxlmem results in p99 latency increases of 9\%, 23\%, and 45\%, respectively, compared to DDR 100\%.
% check
Finally, as expected, allocating 100\% of pages to DDR memory results in the lowest and most stable p99 latency. % until the targeted QPS reaches 100~K. 
% check
Explaining the substantial difference in the p99 latency values for various percentages of pages allocated to \cxlmem and different target QPS values, we note that \texttt{Redis} typically operates with response time at a $\mu$s scale, making it highly sensitive to memory access latency (\S\ref{sec:setup:benchmark}).
% check
Therefore, allocating more pages to \cxlmem and/or increasing the target QPS makes \texttt{Redis} more frequently access \cxlmem with almost 2$\times$ longer latency than DDR memory, resulting in higher p99 latency.
% check
\begin{comment}
Note that \texttt{Redis} begins to drop requests to limit the increase in p99 latency  at high QPS.
% check
To capture the impact of allocating more pages to \cxlmem on both p99 latency and throughput, we also measure the difference between various targeted QPS and delivered QPS. 
% check
This shows that the throughput of \texttt{Redis} saturates at lower QPS as more pages allocated to \cxlmem. 
% check
For example, when allocating 25\%, 50\%, and 75\% of pages to \cxlmem, we observe that the throughput saturates approximately at 104~K, 96~K, and 88~K, respectively.
% check
\end{comment}

\begin{figure}[!t]
    \centering
    %\vspace{-12pt}
    \includegraphics[width=0.7\linewidth]{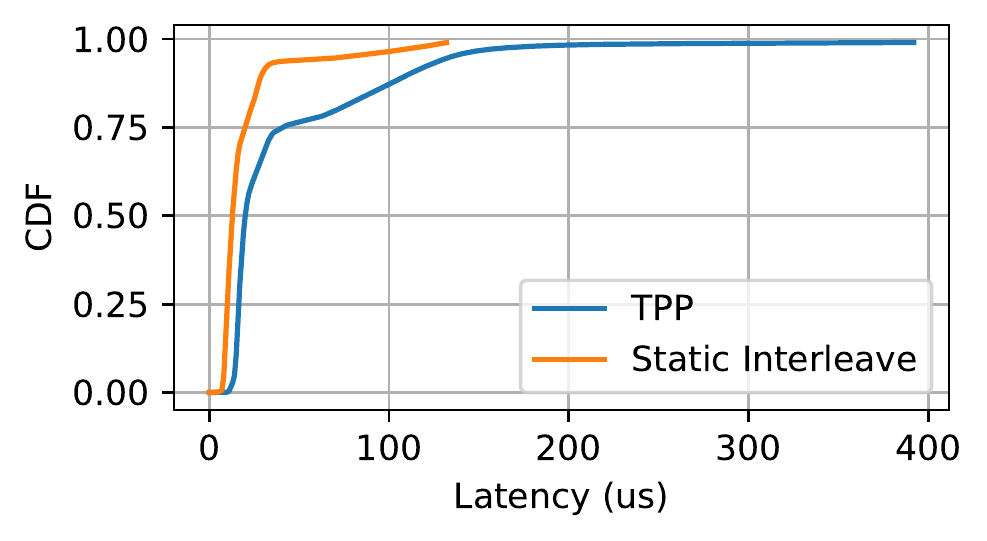}
    \vspace{-12pt}
    \caption{Impact of TPP on latency of  \texttt{Redis} compared with statically allocating 
    %a random static allocation 
    25\% of (random) pages to \cxlmem. We show the distributions up to the p99 latency.}
    \label{fig:app:latency:tpp}
    \vspace{-6pt}
\end{figure}

\niparagraph{Redis+TPP.} We conduct an experiment to assess whether the most recent transparent page placement (TPP)~\cite{tpp} can minimize the impact of using \cxlmem on the p99 latency.
% check
The most recent publicly available release~\cite{tpp-patch} only offers an enhanced page migration policy, and it does not automatically place pages in \cxlmem.
% check
Thus, we begin by allocating 100\% of pages requested by \texttt{Redis} to \cxlmem and let TPP automatically migrate pages to DDR memory until the percentage of the pages allocated to \cxlmem becomes 25\%, based on the DDR to CXL bandwidth ratio in our setup. 
% check
Then we measure the latency values of \texttt{Redis}. 
% check
\figref{fig:app:latency:tpp} compares two distributions of the measured latency values. 
% check
The first one is from using TPP, while the second one is from statically allocating 25\% of pages to \cxlmem.
%and \cxlmem with the 75:25 ratio.
% check
%latency values over time when we use TPP along with p99 latency values from 100\% and 75\% of pages statically allocated to DDR5, respectively.  

TPP migrates a large number of pages to DDR memory in the beginning phase, requiring the CPU to (1) copy pages from one memory device to another and (2) update the OS page table entries associated with the migrated pages~\cite{10.1145/3297858.3304024}. 
Since (1) and (2) incur high overheads, we measure the p99 latency only after 75\% of pages are migrated to DDR memory.
% check
As shown in \figref{fig:app:latency:tpp}, TPP generally gives higher latency, resulting in 174\% higher p99 latency than statically allocating 25\% of pages to \cxlmem. 
% check
This is because TPP constantly migrates a small percentage of pages between DDR memory and \cxlmem over time, based on its metric assessing hotness/coldness of the pages.
% check
Although TPP has a feature that reduces ping-pong behavior (\ie, pages are constantly being promoted and demoted between DDR memory and \cxlmem), migrating pages incurs the overheads from (1) and (2) above. 
% check
(1) blocks the memory controllers from serving urgent memory read requests from latency-sensitive applications~\cite{10.1145/3243176.3243191}, and (2) also requires a considerable number of CPU cycles and memory accesses.
% check

\niparagraph{DSB.} \figref{fig:dsb_compose}--\ref{fig:dsb_mixed} present the p99 latency values of (b) \texttt{compose posts}, (c) \texttt{read user timelines}, and (d) \texttt{mixed workloads}. 
Table~\ref{table:dsb-components} summarizes the components of the benchmarks, their working set sizes and characteristics, and allocated memory devices.
In our experiment, we allocate 100\% of the pages pertinent to the caching and storage components with large working sets 
to either DDR memory (DDR 100\%) or \cxlmem (CXL 100\%). %, because these components can benefit from the capacity expanded by \cxlmem. 
Meanwhile, we always allocate 100\% of the pages associated with the remaining components, such as \texttt{nginx} front-end and analytic docker images (\eg, logic in Table~\ref{table:dsb-components}), to DDR memory, since these components are more sensitive to memory access latency than the caching and storage components. 
For example, \texttt{nginx} spends 60\% of CPU cycles on the CPU front-end, which is dominated by fetching instructions from memory~\cite{deathstartbench}. 
Therefore, %\texttt{nginx} is highly sensitive to memory access latency and 
pages of such components should be allocated to DDR memory. % along with the other memory-latency bounded components. 

\begin{table}[t]
\centering
    %\vspace{-6pt}
    \caption{Components of \texttt{DSB} social network benchmark.}
    \vspace{-8pt}
    \resizebox{\columnwidth}{!}{%
    \begin{tabular}{cccc}
        \toprule
        \textbf{Name}    & \textbf{Working set}  & \textbf{Intensiveness} & \textbf{Allocated mem. type} \\ 
        \midrule
        Frontend & 83~MB & Compute & DDR memory \\
        Logic & 208~MB & Compute & DDR memory \\
        Caching \& Storage & 628~MB & Memory & CXL memory \\
        \bottomrule
    \end{tabular}
    }
    \label{table:dsb-components}
    \vspace{-12pt}
\end{table}

This experiment shows that all three benchmarks, \texttt{compose posts}, \texttt{read user timelines}, and \texttt{mixed workloads} are not sensitive to long latency of accessing \cxlmem as they exhibit little difference in p99 latency values between CXL 100\% and DDR 100\%.
This is because of two reasons.
% check
First, most of the p99 latency in these benchmarks is contributed by the front-end and/or logic components (\ie, DDR 100\%). 
This makes the latency of accessing \cxlmem amortized by the these components,  and thus the p99 latency is much less dependent on the latency of accessing databases (\ie, CXL 100\%).
%such as \texttt{nginx},
%(\ie, \texttt{MongoDB}, \texttt{Redis}, and \texttt{memcached})
% check
Second, the p99 latency of \texttt{DSB} is at a $m$s scale and two orders of magnitude longer than that of \texttt{Redis}. 
% check
Therefore, it is not as sensitive to memory access latency as that of \texttt{Redis}.
% check
%This makes the latency of accessing \cxlmem amortized by the latency-dominant components, such as \texttt{nginx}, and thus the p99 latency is much less dependent on the latency of accessing databases (\ie, \texttt{MongoDB}, \texttt{Redis}, and \texttt{memcached}).
%
%Meanwhile, the p99 latency of \texttt{compos posts} is dominated by accessing databases, making it more sensitive to long latency of accessing \cxlmem. 
%

Note that CXL 100\% provides lower p99 latency values than DDR 100\% for \texttt{mixed workloads} when the QPS range is between 5~K and 11~K.
% check
This is because \texttt{mixed workloads} is far more memory-bandwidth-intensive than \texttt{compose posts} and \texttt{read user timelines}.
% check
Specifically, when we measure the average bandwidth consumption by these three benchmarks in the QPS range that saturates the throughput of the benchmarks, we observe that \texttt{mixed workloads} consumes 32~GB/s while \texttt{compose posts} and \texttt{read user timelines} consume only 7~GB/s and 10~GB/s, respectively.
% check
When a given application consumes such high bandwidth in our setup, we observe that the application's throughput, which is inversely proportional to its latency, becomes sensitive to the bandwidth available for the application (\S\ref{sec:app:throughput}).
% check
Lastly, as the QPS approaches to 11~K, the compute capability of the CPU cores becomes the dominant bottleneck, leading to a decrease in the p99 latency gap between DDR 100\% and CXL 100\%. 
% check
%This analysis on \texttt{DSB} offers a compelling use case for \cxlmem. 
%
%That is, we allocate pages pertinent to latency-sensitive components of a given application to fast DDR memory, 
%while allocating pages associated with latency-insensitive components to slow \cxlmem. 
%

\begin{figure}[t]
    \centering
    %\vspace{-6pt}
    \includegraphics[width=\linewidth]{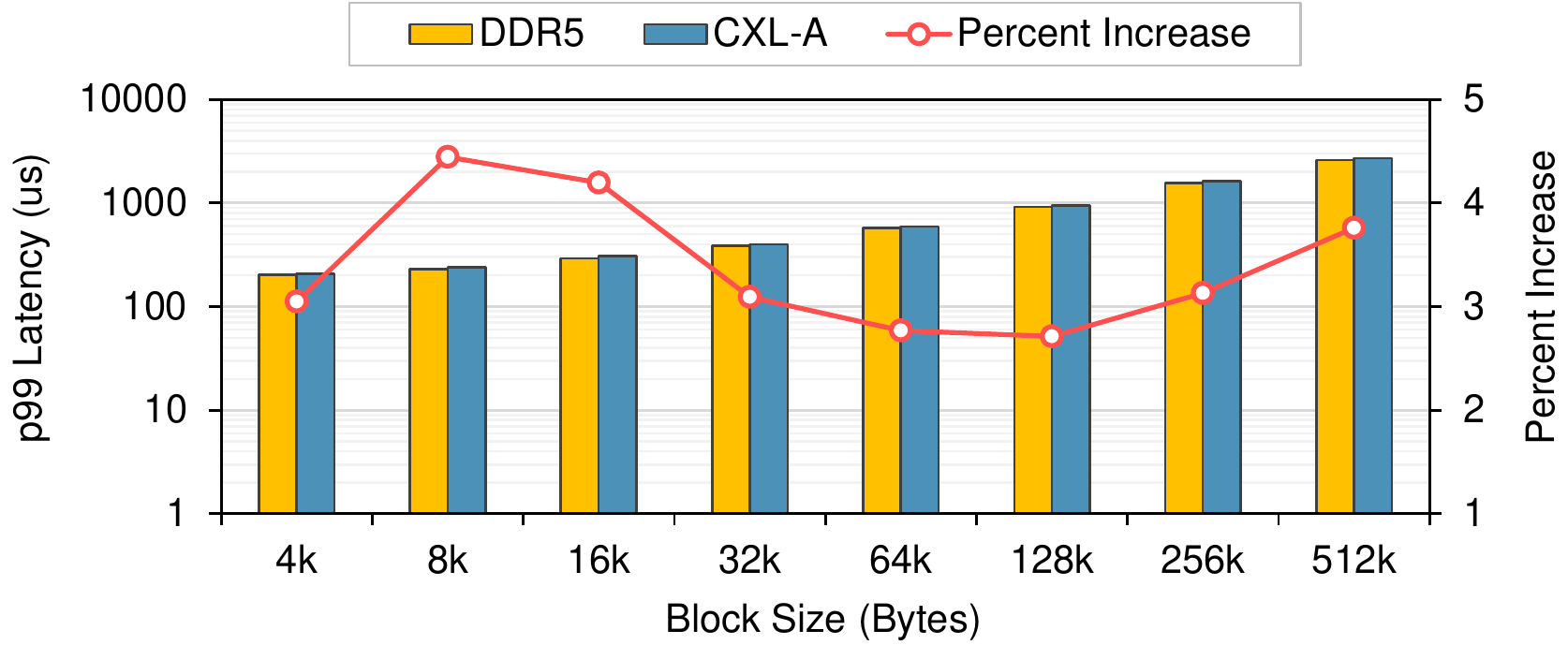}
    \vspace{-15pt}
    \caption{p99 latency values of \texttt{FIO} for various block sizes, and percentage values of increase in p99 latency by allocating page cache to CXL.  
    %We use a \texttt{Zipfian} distribution to evaluate the impact of using page cache on file-system performance.
    } 
    \label{fig:app:latency:fio:log-scale} %\yan{TODO, DDR5-L in the fig}
    \vspace{-9pt}
\end{figure}

\niparagraph{FIO.} 
\figref{fig:app:latency:fio:log-scale} presents the p99 latency values of \texttt{FIO} with 4~GB page cache allocated to either DDR memory or \cxlmem for various I/O block sizes. We use a \texttt{Zipfian} distribution for \texttt{FIO} to evaluate the impact of using page cache on file-system performance.
% check
It shows that allocating the page cache to \cxlmem gives only $\sim$3\% longer p99 latency than DDR5 memory for 4~KB block size. 
% check
This is because the p99 latency for a 4~KB block size is primarily  dominated by the Linux kernel operations related to page cache management, such as context switching, page cache lookup, sending an I/O request through the file system, block layer, and device driver.
% check
However, with a 8~KB block size, the cost of Linux kernel operations is amortized, as multiple 4~KB pages are brought from a storage device by a single system call.
% check
Consequently, longer access latency of \cxlmem affects the p99 latency of \texttt{FIO} more notably, resulting in a $\sim$4.5\% increase in the p99 latency. 
% check
Meanwhile, as the block size increases beyond 8~KB, the page cache hit rate decreases from 76\% for 8~KB to 65\% for 128~KB. 
% check
As lower page cache hit rates necessitate  more page transfers from the storage device, the storage access latency begins to dominate the p99 latency. 
% check
In such a case, the difference in memory access latency between DDR memory and \cxlmem exhibits a lower impact on p99 latency, since Data Direct I/O (DDIO)~\cite{ddio} directly injects pages read from the storage device into the LLC~\cite{alian-ispass,yuan-iat,254372,farshin2019make}. 
% check
Lastly, we observe another trend shift beyond 128~KB block size, which is mainly due to the limited read and write bandwidth of \cxlmem.
% check
As more page cache entries are evicted from the LLC to memory as well as from memory to the storage device, the limited bandwidth of \cxlmem increases the effective latency of I/O requests.
% check

%Overall, this shows that \cxlmem, as a memory capacity expander, can serve as large page cache with little impact on application performance, as prior work~\cite{tpp} suggests.
% check
%Yet, we need to carefully consider the impact of \cxlmem's limited bandwidth on application performance.
% check

\begin{figure*}[!b]
    \centering
    \begin{subfigure}[b]{0.572\textwidth}
        \centering
        \includegraphics[width=\linewidth]{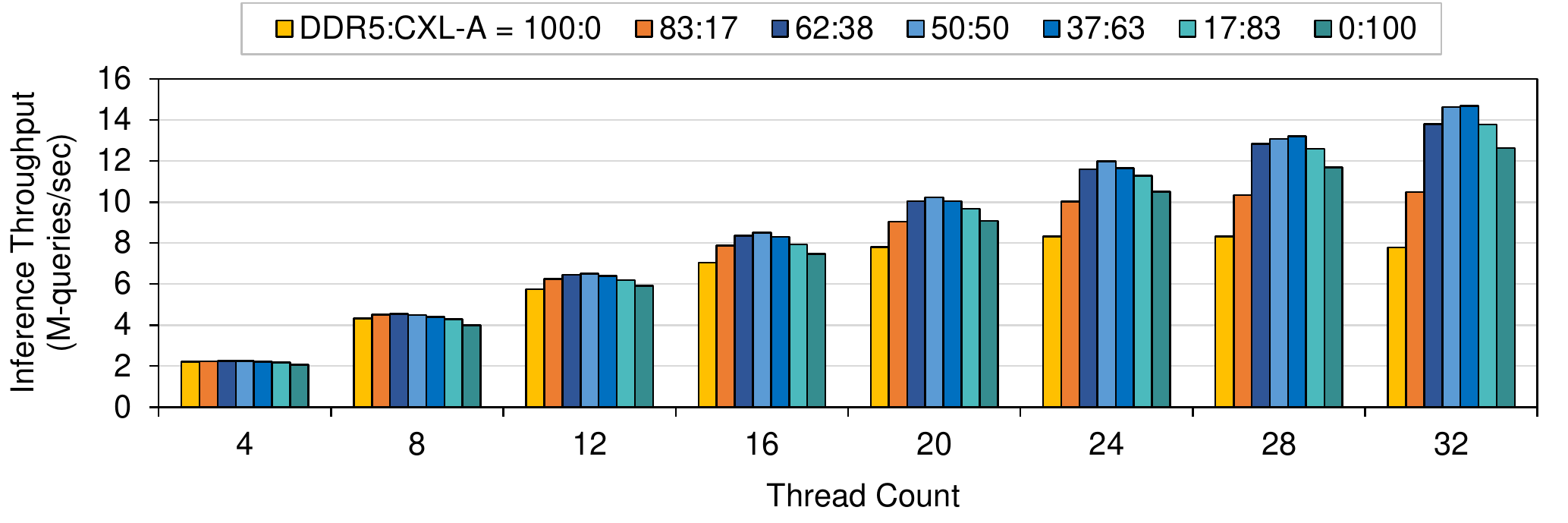}
        \vspace{-15pt}
        \caption{[DLRM] embedding reduction}
        \label{fig:ycsb_max_qps:dlrm}
    \end{subfigure}
    \hfill
    \begin{subfigure}[b]{0.407\textwidth}
        \centering
        \includegraphics[width=\linewidth]{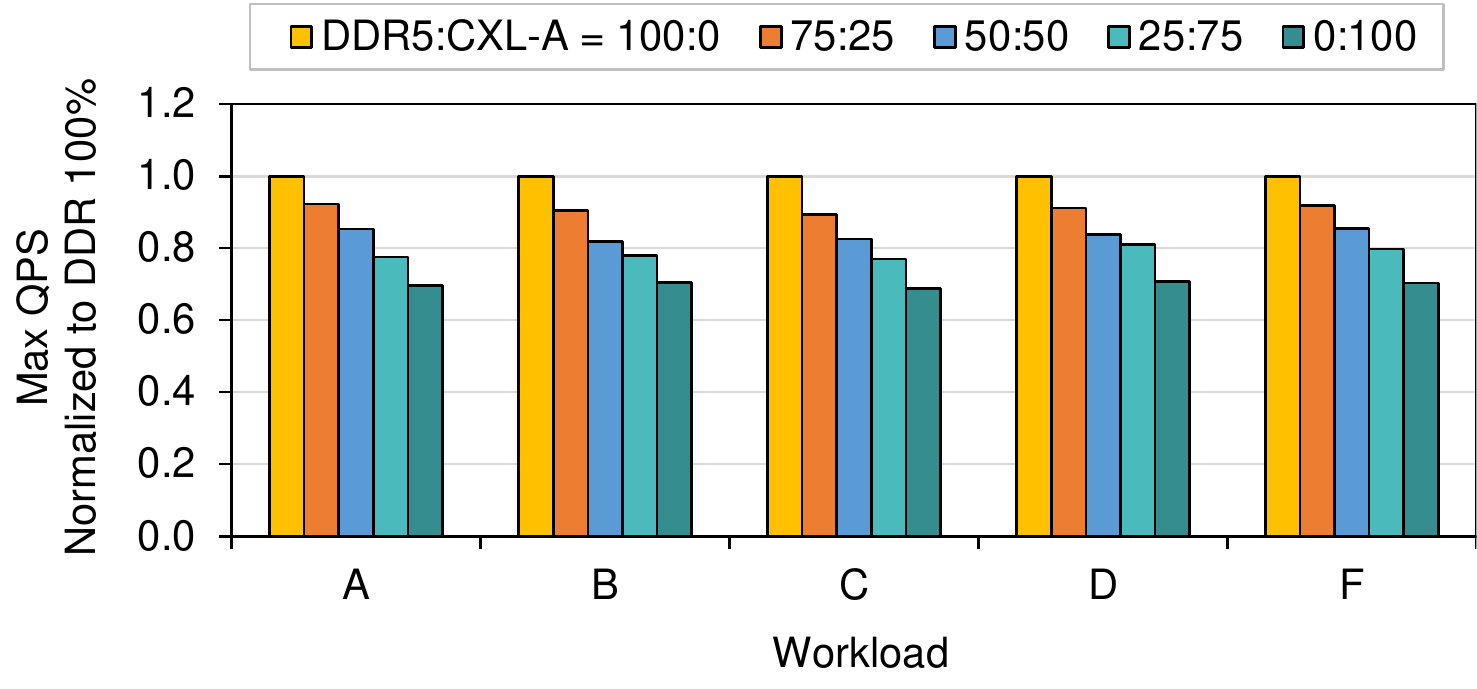}
        \vspace{-15pt}
        \caption{[Redis] YCSB-A}
        \label{fig:ycsb_max_qps:redis}
    \end{subfigure}
    \hfill
    \vspace{-6pt}
    \caption{Impact of using \cxlmem on throughput of \texttt{Redis} and \texttt{DLRM} for various ratios of page allocation to \cxlmem.
    %for various CXL memory configurations. The legend denotes percent of Redis memory allocated to CXL memory. \texttt{YCSB} workload \texttt{D} defaults to read the most recently inserted elements \texttt{D-lat}, but we also evaluate this workload with read request in Zipfian \texttt{D-zipf} or uniform \texttt{D-uni} distribution to see the effect on access locality.
    }
    \label{fig:ycsb_max_qps}
    \vspace{-6pt}
\end{figure*}

\niparagraph{Key findings.} 
Based on our analyses above, we present the following three key \textbf{\underline{F}}indings. 
% check
\textbf{(F1)} Allocating any percentage of pages to \cxlmem proportionally increases the p99 latency of simple memory-intensive applications demanding $\mu$s-scale latency, since such applications are highly sensitive to memory access latency. %, especially when considering the p99 latency.
\textbf{(F2)} Even an intelligent page migration policy may further increase the p99 latency of such latency-sensitive applications because of the overheads of migrating pages. 
% check
\textbf{(F3)} Judiciously allocating certain pages to \cxlmem does not increase the p99 latency of complex applications exhibiting $m$s-scale latency. 
This is because the long latency of accessing \cxlmem marginally contributes to the end-to-end latency of such applications and it is amortized by
intermediate operations between accesses to \cxlmem.
%\textbf{(F4)} For memory-bandwidth-intensive complex applications, allocating n pages
%can be hidden (or amortized) by 
% experience practically no increase in p99 latency even when allocating 100\% of pages of the storage and cache components to \cxlmem. 
% check
%This is 

\subsection{Throughput}
\label{sec:app:throughput}
\niparagraph{DLRM.} \figref{fig:ycsb_max_qps:dlrm} shows the throughput of \texttt{DLRM} embedding reduction for various percentages of pages allocated to \cxlmem.
% check
As the throughput of \texttt{DLRM} embedding reduction is bounded by memory bandwidth~\cite{10.1109/ISCA45697.2020.00070,asplos21lee,280856}, it begins to saturate over 20 threads when 100\% of pages are allocated to DDR memory. 
% check
In such a case, we observe that allocating a certain percentage of pages to \cxlmem can improve the throughput further, as it supplements to the bandwidth of DDR memory, increasing the total bandwidth available for \texttt{DLRM}.
% check
For instance, when running 32 threads, we observe that allocating 63\% of pages to \cxlmem can maximize the throughput of \texttt{DLRM} embedding reduction, providing 88\% higher throughput than DDR 100\%.
% check
Note that a lower percentage of pages will be allocated to \cxlmem for achieving the maximum throughput if the maximum bandwidth capability of a given \cxlmem device is lower (\eg, CXL-C). 
% check
This clearly demonstrates the benefit of \cxlmem as a memory bandwidth expander. 
% check
%We have also observed that allocating certain pages of \texttt{DSB} \texttt{mixed workload}  to \cxlmem reduces the p99 latency (\S\ref{sec:app:latency})

\niparagraph{Redis.} Although \texttt{Redis} is a latency-sensitive application, its throughput is also an important performance metric.
% check
\figref{fig:ycsb_max_qps:redis} shows the maximum sustainable QPS for various percentages of pages allocated to \cxlmem.
% check
For example, for \texttt{YCSB-A}, allocating 25\%, 50\%, 75\%, and 100\% of pages to \cxlmem provides 8\%, 15\%, 22\%, and 30\% lower throughput than allocating 100\% of pages to DDR memory.  
As \texttt{Redis} does not fully utilize the memory bandwidth, its throughput is bounded by memory access latency.
% check
Thus, similar to its p99 latency trends (\figref{fig:ycsb_redis}), allocating more pages to \cxlmem reduces the throughput of \texttt{Redis}.
%as interleaving with \cxlmem will always introduce higher memory access latency.
% check
%
%
%
\begin{comment}
Note that \texttt{Redis} begins to drop requests to limit the increase in p99 latency  at high QPS.
% check
To capture the impact of allocating more pages to \cxlmem on both p99 latency and throughput, we also measure the difference between various targeted QPS and delivered QPS. 
% check
This shows that the throughput of \texttt{Redis} saturates at lower QPS as more pages allocated to \cxlmem. 
% check
For example, when allocating 25\%, 50\%, and 75\% of pages to \cxlmem, we observe that the throughput saturates approximately at 104~K, 96~K, and 88~K, respectively.
% check
\end{comment}

\niparagraph{Key findings.} 
Based on our analyses above, we present the following %two 
key \textbf{\underline{F}}inding. 
% check
\textbf{(F4)} For memory-bandwidth-intensive applications, na\"ively allocating 50\% of pages to \cxlmem based on the OS default policy may result in lower throughput than allocating 100\% of pages to DDR memory, even with higher total bandwidth from using both DDR memory and \cxlmem together.
% check
This motivates us to develop a dynamic page allocation policy that can automatically configure the percentage of pages allocated to \cxlmem at runtime based on the bandwidth capability of a given CXL memory device and bandwidth consumed by co-running applications (\S\ref{sec:policy}).
% check

\begin{table}[t!]
\centering
%\vspace{-6pt}
\caption{Throughput of \texttt{DLRM} using only 1 SNC node versus all 4 SNC nodes, normalized to the throughput of \texttt{DLRM} running on 1 SNC node allocating all the pages to local DDR memory.}
\vspace{-6pt}
\label{tab:throughput_snc}
\begin{center}
\resizebox{0.7\columnwidth}{!}{%
\begin{tabular}{cccc}
\hline
\multicolumn{2}{c}{\textbf{1 SNC node}}  & \multicolumn{2}{c}{\textbf{4 SNC nodes}}\TBstrut \\ 
\cmidrule(lr){1-2}\cmidrule(lr){3-4}
\textbf{DDR 100\%} & \textbf{CXL 100\%} & \textbf{DDR 100\%} & \textbf{CXL 100\%}\Bstrut \\
\hline
1 & 0.947 & 1 & 0.504 \TBstrut \\
\hline
\end{tabular}
}
\label{tab2}
\end{center}
\vspace{-6pt}
\end{table}

\subsection{Interaction with Cache Hierarchy}
\label{sec:app:cache}
Previously, we discussed that accessing \cxlmem breaks the LLC isolation among SNC nodes (\S\ref{sec:mem-charac:cache}). 
% check
To analyze the impact of such an attribute of accessing \cxlmem on application performance, we evaluate two cases. 
In the first case (`1 SNC node' in Table~\ref{tab:throughput_snc}), only one SNC node (\ie, SNC-0 in \figref{fig:spr_interation_cache_hierarchy}) runs 8 \texttt{DLRM} threads while the other three SNC nodes idle. 
% check
In the second case (`4 SNC nodes' in Table~\ref{tab:throughput_snc}),
%\figref{fig:app:latency:dlrm:interference}), 
%the four SNC nodes run 32 \texttt{DLRM} threads, with 
each SNC node runs 8 \texttt{DLRM} threads. 
% check
Only SNC-0 allocates 100\% of its pages to either its DDR memory or \cxlmem, while the other three SNC nodes (\ie, SNC-1, SNC-2, and SNC-3) allocate 100\% of their pages only to their respective local DDR memory. 
% check
The second case is introduced to induce interference at the LLC among all the SNC nodes when SNC-0 with CXL 100\%.
%
%Specifically, 
%cache lines in the L2 caches of SNC-1, SNC-2, and SNC-3 are always evicted only to their respective LLC slices. 
%
%Since cache lines in the L2 caches of SNC-0 with CXL 100\% are also evicted to LLC slices coupled with SNC-1, SNC-2, and SNC-3, the effective LLC capacity of SNC-1, SNC-2, and SNC-3 is reduced. 
%
%On the other hand, cache lines in the L2 caches of SNC-0 with CXL 100\% can be evicted to LLC slices in any SNC nodes. 
%
%As a result, the effective LLC capacity of SNC-0 would be reduced compared to the first case, because the LLC slices of SNC-1, SNC-2, and SNC-3 are now shared between L2 cache lines evicted from SNC-0 and those from SNC-1, SNC-2, and SNC-3, respectively.
%
%At the same time, SNC-0 perceives that SNC-1, SNC-2, and SNC-3 reduce its effective LLC capacity (\ie, LLC slices from all the SNC nodes) since they also evict LLC lines within their respective LLC slices and some of the evicted LLC lines are from the L2 caches of SNC-0.
% chck

Table~\ref{tab:throughput_snc} shows that SNC-0 with CXL 100\% in `1 SNC node' offers 88\% higher throughput than SNC-0 with CXL 100\% in `4 SNC nodes.'
This is because of the other three SNC nodes in `4 SNC nodes' reduces the effective LLC capacity of SNC-0 with CXL 100\% (\ie, LLC slices from all the SNC nodes).
Specifically, while the other three SNC nodes evict LLC lines within their respective LLC slices, they also inevitably evict many LLC lines from SNC-0 with CXL 100\%.
% check
Although not shown in Table~\ref{tab:throughput_snc}, SNC-0 with DDR 100\% in `1 SNC node'  provides 2\% higher throughput than each of the other three SNC nodes in `4 SNC nodes' when SNC-0 in `4 SNC nodes' is with CXL 100\%.
% check
This is because SNC-0 with CXL 100\% in `4 SNC nodes' pollutes the LLC slices of the other three SNC nodes with cache lines evicted from the L2 caches of SNC-0, breaking the LLC isolation among the SNC nodes. 
% check
Lastly, in our previous evaluation of \texttt{DLRM} throughput (\S\ref{sec:app:throughput}), when SNC-0 needs to run more than 8 threads of \texttt{DLRM} in the SNC mode, it makes the remaining threads run on the CPU cores in the other three SNC nodes.
% check
Nonetheless, the CPU cores in the other three SNC nodes continue to access the DDR memory of SNC-0, and cache lines in the L2 caches of these CPU cores are still evicted to the LLC slices of SNC-0 since the cache lines were from the DDR memory of SNC-0.
% check

% %\input{6-advise}

\section{CXL-Memory-aware Dynamic \\ Page Allocation Policy}
\label{sec:policy}
%We have demonstrated that \cxlmem can also serve as a bandwidth expander to improve the performance of bandwidth-intensive applications (\S\ref{sec:app:throughput}).
We have demonstrated a potential of \cxlmem as a bandwidth expander, which can improve the performance of bandwidth-intensive applications (\S\ref{sec:app:throughput}).
% check
If the throughput of a given application is limited by the bandwidth, allocating a higher percentage of pages to \cxlmem may alleviate bandwidth pressure on DDR memory, and hence reduce average memory access latency.
% check
Intuitively, such a percentage should be tuned for different \cxlmem devices given their distinct bandwidth capabilities (\S\ref{sec:app:throughput}).
% check
By contrast, if a given application is not memory-bandwidth-bounded, allocating a lower percentage of pages to \cxlmem may lead to lower average memory access latency and thus higher throughput.
% check
That stated, to better utilize auxiliary memory bandwidth provided by \cxlmem, we present \policy, a dynamic page allocation policy.
% check
\policy automatically tunes the percentage of new pages to be allocated by the OS to \cxlmem based on three factors: (1) bandwidth capability of \cxlmem, (2) memory intensiveness of co-running applications, and (3) average memory access latency.
% check
Note that \policy, focusing on the page allocation ratio between DDR memory and \cxlmem, is orthogonal and complementary to \texttt{TPP}. 
% check

\begin{figure} % [b]
    \centering
%    \vspace{-9pt}
    \includegraphics[width=0.92\linewidth]{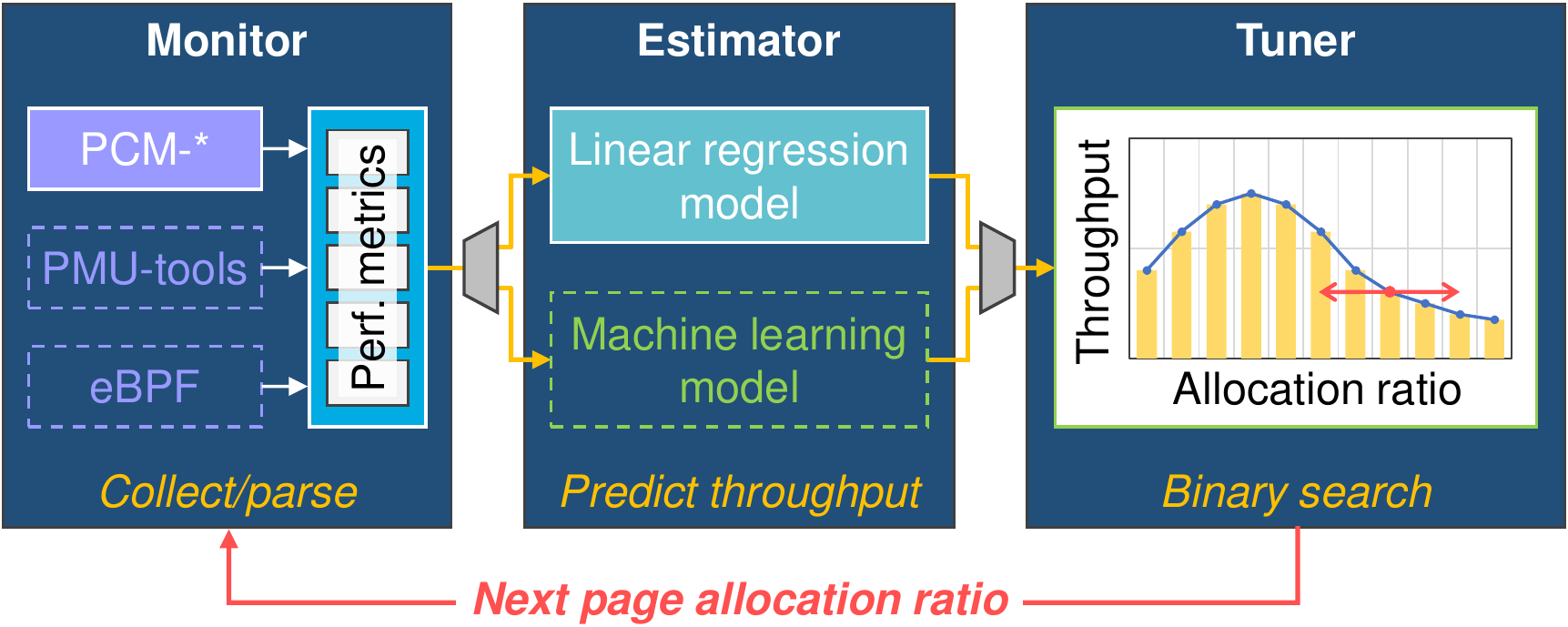}
    \vspace{-3pt}
    \caption{Overview of \policy. The components in dotted boxes can be used for better performance.}
    \label{fig:policy-overview}
    \vspace{-6pt}
\end{figure}

\subsection{Policy Design}
\label{sec:policy:design}
\policy consists of three runtime \textbf{\underline{M}}odules (\figref{fig:policy-overview}).
% check
\textbf{(M1)} periodically monitors some CPU counters related to memory subsystem performance, and then \textbf{(M2)} estimates memory-subsystem performance based on values of the counters. 
% check
When a given application requests an allocation of new pages, (M3) tunes the percentage of the new pages allocated to \cxlmem, aiming to improve the throughput of the application. 
% check
Subsequently, \texttt{mempolicy}~\cite{mn_il_patch} sets the page allocation ratio between DDR memory and \cxlmem based on the percentage guided by (M3), and instructs the OS to allocate the new pages based on the ratio.  
% check

\begin{table}[b]
\centering
    \vspace{-6pt}
    \caption{CPU counters pertinent to memory-subsystem perf.}
    \vspace{-6pt}
    \resizebox{\columnwidth}{!}{%
    \begin{tabular}{ccc}
        \toprule
        \textbf{Metric} & \textbf{Tool} & \textbf{Description} \\ 
        \midrule
        L1 miss latency & pcm-latency & Average L1 miss latency (ns)\\
        DDR read latency & pcm-latency & DDR read latency (ns)\\
        IPC & pcm & Instructions per cycle \\
        \bottomrule
    \end{tabular}
    }
    \label{table:model-metric}
\end{table}

\begin{figure}[t!]
    \centering
    \begin{subfigure}[b]{0.517\columnwidth}
        \includegraphics[width=\linewidth]{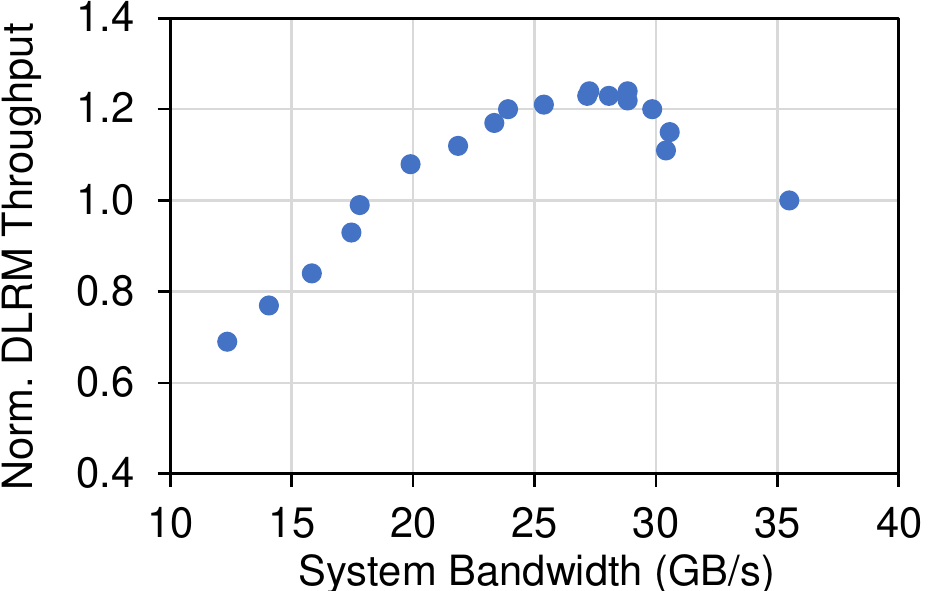}
        \vspace{-12pt}
        %\caption{Throughput vs. System BW}
        \caption{System bandwidth}
        \label{fig:correlation:bandwidth}
    \end{subfigure}
    \hfill
    \begin{subfigure}[b]{0.463\columnwidth}
        \includegraphics[width=\linewidth]{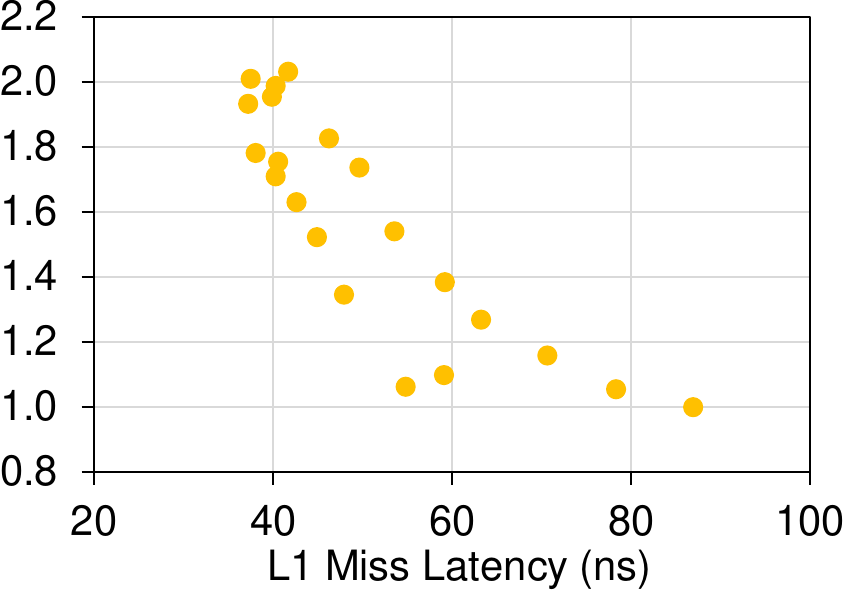}
        \vspace{-12pt}
        %\caption{Throughput vs. L1 miss latency}
        \caption{L1 miss latency}
        \label{fig:correlation:latency}
    \end{subfigure}
    \vspace{-3pt}
    \caption{Correlations between throughput and various counter values, as we increase the percentage of pages allocated to \cxlmem for \texttt{DLRM}. 
    The system bandwidth is 
    %the total memory bandwidth provided by both DDR and \cxlmem, and 
    %normalized to 
    the total consumed memory bandwidth, and
    %given by only DDR. 
    The throughput is normalized to DDR 100\%.%the case allocating all the pages to DDR. 
    %\nskim{for x-axis, we can show the absolute values? For System BW, you can put a dotted vertical line, and label as DDR max bandwidth. BTW, this plot can be taller and use only one y-axis label as both plots shows DLRM throughput.}
    }
    \label{fig:correlation}
    \vspace{-6pt}
\end{figure}

\niparagraph{(M1) Monitoring CPU counters related to memory subsystem performance.} 
We use Intel PCM~\cite{intel-pcm} % and PMU~\cite{pmu-tools} 
to periodically sample various CPU counters related to memory subsystem performance, as listed in Table~\ref{table:model-metric}.
% check
These CPU counters allow (M2) to estimate overall memory-subsystem performance.
% check
In \figref{fig:correlation}, we run \texttt{DLRM}, of which the throughput is bounded by memory bandwidth.
% check
Then we observe correlations between \texttt{DLRM} throughput and values of those counters, as we vary the percentage of pages allocated to \cxlmem.
% check

\figref{fig:correlation:bandwidth} shows that \texttt{DLRM} throughput is proportional to the consumed memory bandwidth.
% check
Yet, as the consumed memory bandwidth exceeds a certain amount, the memory access latency rapidly increases due to contention and resulting queuing delay at the memory controller~\cite{tootoonchian2018resq}, which, in turn, decreases the \texttt{DLRM} throughput.
% check

Meanwhile, \figref{fig:correlation:latency} shows that
\texttt{DLRM} throughput is inversely proportional to L1 miss latency.
% check
The L1 miss latency is an important memory-subsystem performance metric that simultaneously captures both the cache friendliness and bandwidth intensiveness (\ie, queuing delay at the memory controller) of given (co-running) applications at the same time. 
% check
At first, allocating more pages to \cxlmem reduces pressure on the DDR memory controller, thereby decreasing the latency of accessing DDR memory and handling L1 misses.
% check
However, at some point, the long latency of accessing \cxlmem begins to dominate that of handling L1 misses, and the application throughput begins to decrease.
% check
Finally, IPC is another important metric that implicitly measures the efficiency of the memory subsystem for the applications. 
% check

\niparagraph{(M2) Estimating system throughput.} To build a model that estimates the system performance, we collect CPU counter values at various DDR:CXL ratios while running \texttt{DLRM} with 24 threads. 
% check
We then build a linear-regression model that correlates these counter values with \texttt{DLRM} throughput. 
% check
Taking these counter values from (M1), \policy periodically estimates (or infers) memory-subsystem performance at runtime. 
% check
In our current implementation of \policy, (M1) samples the counters every 1 second. To reduce the noise among the values, we collect a moving average of the past 5 samples for each counter. The averaged value is then fed into (M2) for performance estimation.
% check
Although we may use a machine-learning (ML) model, we use the following simple linear model for the current implementation of \policy:
% check

\begin{equation}\label{eq:2}
%\begin{aligned}
Y = \beta_{0} + \beta_1 X_1 + \beta_2 X_2 + ...
%\end{aligned}
\end{equation}

\noindent where $Y$ represents the estimated memory-subsystem performance, $X_n$ represents a counter value listed in Table~\ref{table:model-metric}, and $\beta_n$ represents the $X_n$'s weight obtained through multiple linear regression steps. 
% check
This linear model is simple enough to be used by the OS at a low performance cost, yet effective enough to estimate the memory-subsystem performance. %, which is proportional the throughput of memory-bandwidth-intensive applications. 
% check
In the current implementation of \policy, we find that using \texttt{PCM} toolkit is sufficient.
% check
Nonetheless, we may use PMU tools and eBPF~\cite{ebpf} to access more diverse counters, facilitating a more precise estimation of memory-subsystem performance.
% check

\SetKwComment{Comment}{/* }{ */}
\RestyleAlgo{ruled}

\setlength{\textfloatsep}{10pt}% Modify \textfloatsep
\setlength{\textfloatsep}{10pt}% Modify \textfloatsep
\begin{algorithm}[t]
\DontPrintSemicolon
\caption{\policy tuning algorithm. \texttt{state}, \texttt{step} and \texttt{ratio} represent memory subsystem performance, unit of tuning page allocation ratio, and ratio of page allocation between DDR and \cxlmem.} \label{alg:tune}
%\tcp*[l]{take reverse direction}
    \While {$true$} {
        $curr\_state\gets estimator()$\;
        \If{$curr\_state < prev\_state$} {
            $curr\_step\gets prev\_step \times (-0.5)$ \tcp*[l]{reverse}
        } 
        $curr\_ratio\gets prev\_ratio + curr\_step$\;
        $check\_ratio\_bound()$\;
        $set\_ratio(curr\_ratio)$\;
        \If{new allocations} {
            $prev\_state\gets curr\_state$\;
            $prev\_step\gets curr\_step$\;
            $prev\_ratio\gets curr\_ratio$\;
        }
        $sleep(tune\_interval)$
    }
\end{algorithm}
% \begin{algorithm}[t]
%     \DontPrintSemicolon
%     \caption{\policy tuning algorithm. \texttt{c}, \texttt{s} and \texttt{r} represent memory-subsystem performance, unit of tuning page allocation, and ratio of page allocation between DDR and \cxlmem. \texttt{H\textsubscript{reset}} and \texttt{H\textsubscript{idle}} are empirically determined thresholds for tuning.} \label{alg:tune}
%     \While {$true$} {
%         $c$ append(Estimator())\;
%         \If{new allocations} {
%             $c_{avg1} = $ average($c$)\;
%             $\Delta = c_{avg1} - c_{avg0}$\;
%             \If{abs($\Delta$) $< H_{idle}$} {
%                 continue\;
%             }
%             \If{abs($\Delta$) $> H_{reset}$} {
%                 reset\_ratio()\;
%                 continue 
%             }
%             \ElseIf{$\Delta < 0$} {
%                 $s_1 = s_0 \times - \frac{1}{2}$ \tcp*[l]{reverse}

%             }
%             $r_1 = r_0 $ + $ s_1 $\;
%             $r_1 = $ check\_ratio\_bound($r_1$)\;
%             set\_ratio($r_1$)\;
            
%             $r_0 = r_1$\;
%             $s_0 = s_1$\;
%             $c_{avg0} = c_{avg1}$\;
%             clear $c$
%         }
%         sleep(tune\_interval)\;
%     }
% \end{algorithm}

\niparagraph{(M3) Tuning the percentage of pages allocated to CXL memory.} 
When a given application demands allocation of memory pages, \policy (Algorithm~\ref{alg:tune}) first compares the estimated memory-subsystem performance value from the past period (line: 9--11) with the current period (line: 3).
% check
If the memory-subsystem performance in the current period has increased compared to the previous period, \policy assumes that its previous decision, \ie, increasing (or decreasing) the percentage of pages allocated to \cxlmem, was correct.
% check
Then it continues to incrementally increase (or decrease) the percentage by a fixed amount (line: 5). 
Otherwise, it will begin to reverse the step by half (line: 4), which decreases (or increases) the percentage and evaluate the decision in the future period to determine a favorable percentage of pages allocated to \cxlmem. Note that the absolute value of the step variable has the minimum limit (\eg, 9\% in our evaluation) to prevent it from being close to zero.
% check
Lastly, inspired by conventional control theory, \policy implements mechanisms to efficiently handle very small or sudden large changes in memory subsystem performance, even though they are not described in Algorithm~\ref{alg:tune}.
% check

%\vspace{-0.05in}
\subsection{Evaluation}
\label{sec:policy:eval}
We have developed \policy after analyzing the various performance characteristics and memory subsystem statistics of \texttt{DLRM}.
% check
However, we expect that \policy should work well for other applications because the monitored L1 miss latency, DDR read latency, and IPC counters are fundamental memory subsystem performance metrics that are strongly correlated with the throughput of memory-bandwidth-intensive applications; 
% check
we believe CXL memory access latency and bandwidth statistics are also useful for estimating memory subsystem performance, but we currently cannot access the corresponding counters.
% check
To demonstrate this, we evaluate the efficacy of \policy by co-running (1) \texttt{SPEC-Mix}, various mixes of memory-intensive SPECrate CPU2017 benchmarks, and (2) \texttt{Redis} and \texttt{DLRM} without measuring their performance characteristics and memory-subsystem statistics in advance.
% check

\begin{figure} %[!b]
    \centering
    %\vspace{-12pt}
    \begin{subfigure}[b]{\linewidth}
        \centering
        \includegraphics[width=\linewidth]{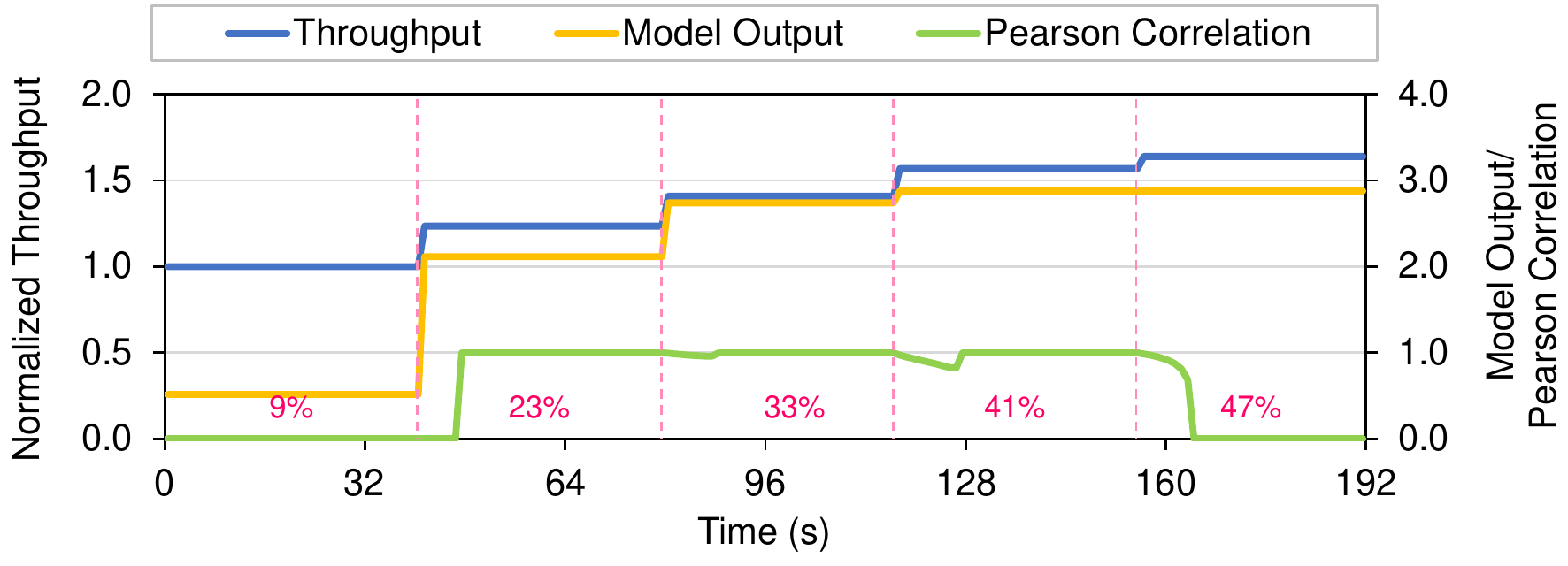}
        \vspace{-12pt}
        \caption{DLRM}
        \vspace{6pt}
        \label{fig:linear-model-throughput:dlrm}
    \end{subfigure}
    \hfill
    \begin{subfigure}[b]{\linewidth}
        \centering
        \includegraphics[width=\linewidth]{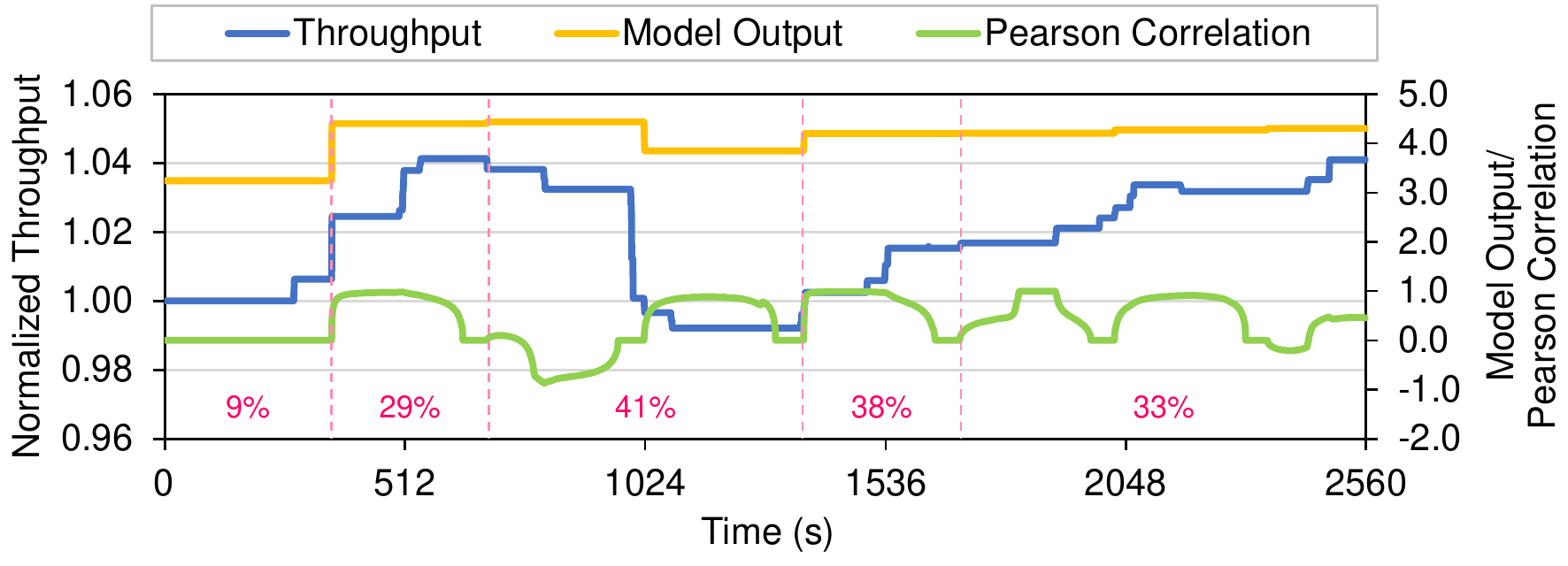}
        \vspace{-12pt}
        \caption{SPEC-Mix}
        \label{linear-model-throughput:spec}
    \end{subfigure}
    \hfill
    \vspace{-18pt}
    \caption{Estimated memory-subsystem performance, measured application throughput, and Pearson coefficient values over time. 
    %For \texttt{DLRM} we simply sweep the percentage of pages allocated to \cxlmem. 
    %For \texttt{SPEC-Mix} we let \policy automatically tune the percentage whenever a benchmark is launched. 
    The numbers represent the percentage values of pages allocated to \cxlmem.
    }
    %after completing the execution of the previous one.} 
    %We use the total instruction per second (IPS) of all the co-running benchmark programs as throughput for \texttt{SPEC-Mix}. We take the throughput value of the first 1~s period to normalize throughput values over time. We sample CPU counters every 1~s which limited by the minimum sampling period allowed by the PCM tool.    }
    \label{fig:linear-model-throughput}
    \vspace{-6pt}
\end{figure}

\figref{fig:linear-model-throughput} shows normalized measured throughput (\texttt{Throughput}), normalized estimated memory-subsystem performance (Eq. (\ref{eq:2}), \texttt{Model Output}), and Pearson correlation coefficient values. %, as we vary the percentage of pages allocated to \cxlmem over time.
For \texttt{DLRM} we simply sweep the percentage of pages allocated to \cxlmem over time.
% check
For \texttt{SPEC-Mix}, we let \policy automatically tune the percentage of pages allocated to \cxlmem whenever a benchmark completes its execution.
% check
The Pearson correlation method allows us to quantify synchrony between time-series data~\cite{benesty2009pearson}.
% check
The coefficient value can range from -1 and 1, indicating that both sets of data trend the same direction when it is positive.
% check
We calculate the Pearson correlation coefficient values to assess the effectiveness of the  estimation model, as Algorithm~\ref{alg:tune} depends on precisely determining only the direction of performance changes after tuning the percentage of pages allocated to \cxlmem.
\figref{fig:linear-model-throughput} demonstrates that the Pearson correlation coefficient values mostly remains positive for both \texttt{DLRM} and \texttt{SPEC-Mix}. 
% check
This indicates that the estimation model is adequate  for Algorithm~\ref{alg:tune} to effectively tune both \texttt{DLRM} and \texttt{SPEC-Mix}. 
% check
It is important to note that the estimation model is based on the weight values derived by fitting counter values from \texttt{DLRM} exclusively in this paper. 
% check
However, it has the potential for further improvement by fitting counter values from a more diverse range of applications.
% check

\begin{figure}[t]
    \centering
    %\vspace{10pt}
    \includegraphics[width=\linewidth]{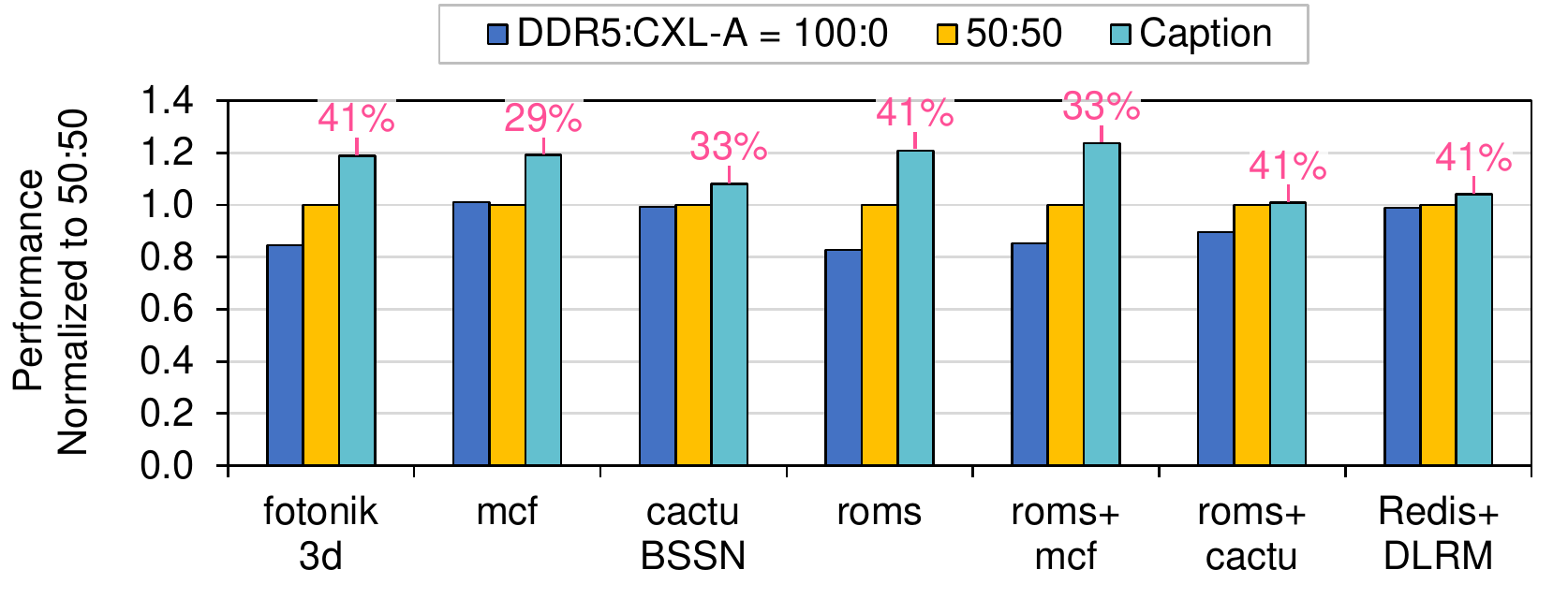}
    \vspace{-16pt}
    \caption{Throughput of each evaluated benchmark or mix, normalized to that with the default static policy allocating 50\% of pages to \cxlmem. A number atop each bar is the percentage of pages allocated to \cxlmem by \policy.}
    \label{fig:tune-improvement}
    \vspace{-6pt}
\end{figure}

\figref{fig:tune-improvement} evaluates the efficacy of \policy for 16 instances of individual \texttt{SPEC} benchmarks, two different \texttt{SPEC-Mix}, and a mix of \texttt{Redis} and \texttt{DLRM}.
% check
For all the evaluation cases, \policy outperforms both 100\% and 50\% allocations to DDR memory while allocating substantial percentages of pages to \cxlmem.  
% check
Specifically, \policy offers 19\%, 18\%, 8\%, and 20\% higher throughput values for \texttt{fotonik3d}, \texttt{mcf}, \texttt{roms}, and \texttt{cactuBSSN}, respectively, than the best static allocation policy (\ie, 100\% or 50\% allocation to DDR memory), allocating 29\%--41\% of pages to \cxlmem in a steady state.
% check
For the mixes of \texttt{mcf} and \texttt{roms}, \texttt{cactuBSSN} and \texttt{roms}, and  \texttt{Redis} and \texttt{DLRM}, \policy provides 24\%, 1\%, and 4\% higher throughput values than the best static allocation policy, allocating 33\%--41\% of pages to \cxlmem.
Since \texttt{DLRM} and \texttt{Redis} use different throughput metrics, we show a geometric mean value of normalized throughput values of \texttt{DLRM} and \texttt{Redis} as a single throughput value. 
These demonstrate that \policy captures the immense memory pressure from co-running applications and tunes to the percentage of pages to \cxlmem that yields higher throughput than the static allocation policies.

In \figref{fig:tune-improvement}, we do not compare \policy with the static allocation policy for \texttt{DLRM} and \texttt{Redis} individually.
This is because we have derived the estimation model after running \texttt{DLRM} (\S\ref{sec:policy:design}) and demonstrated that allocating all the pages to DDR memory is best for \texttt{Redis} (\S\ref{sec:app:throughput}).
However, our evaluation shows that \policy presents 80\% higher and 4\% lower throughput values than allocating 100\% and 50\% of pages to DDR memory, respectively, for \texttt{DLRM}.
% check
For \texttt{Redis}, \policy is able to identify that allocating more memory to low-latency DDR memory is beneficial, and thus offers 3.2\% higher throughput than allocating 50\% of pages to DDR memory but 8.6\% lower throughput than allocating 100\% of pages to DDR memory.
Albeit not perfect, \policy demonstrates its capability of searching near-optimal percentage values of pages allocated to \cxlmem without any guidance from users and/or applications for several workloads with notably different characteristics.
Lastly, one of our primary goals is to emphasize the need for a dynamic page allocation policy and show a potential of such a policy. Hence, we leave further enhancement of \policy as future work.

%This improvement is \textcolor{red}{X\%} smaller than the maximum improvement that we observed in \figref{fig:ycsb_max_qps}. but it demonstrates that. 
% !TEX root = paper.tex
\section{Related Work}
\label{sec:related}
With the rapid development of memory technologies, diverse heterogeneous memory devices have been introduced.
These memory devices are often different from the standard DDR-based DRAM devices, and each memory device offers unique characteristics and trade-offs. 
These include but are not limited to persistent memory, such as Intel Optane DIMM~\cite{9251957,fast20yang,eurosys22xiang}, remote/disaggregated memory~\cite{farm, infiniswap, socc20kalia,10.1145/1555754.1555789,10.1145/3520263.3534650,10.1145/3387514.3405897,10.1145/3357526.3357543}, and even byte-addressable SSD~\cite{10.1145/3297858.3304061,8416845}. 
These heterogeneous memory devices in the memory hierarchy of datacenters have been applied to diverse domains of applications. 
For example, in a tiered memory/file system, pages can be dynamically placed, cached, and migrated across different memory devices, based on their hotness and persistency requirements~\cite{tpp,hemem,227786,285754,10.1145/3582016.3582031,258860,10.1145/3445814.3446745,8988604,10.1145/3297858.3304024,273808,10.1145/3079856.3080245,10.1145/3037697.3037706,8676386,6212453,10.1145/3445814.3446713,9065506,10.1145/3410463.3414672,8327039}. 
Besides, database or key-value store can leverage these memory devices for faster and more scalable data organization and retrieval~\cite{slmdb,novelsm,listdb,matrixkv,ChameleonDB,viper,flatstore,utree,dptree}.
Solutions similar to \policy were proposed in the context of HBM~\cite{memsys17-batman} and storage system ~\cite{fast21-wu}.
While they have been extensively profiled and studied, \cxlmem, as a new member in the memory tier, still has unclear performance characteristics and indications, especially its interaction with CPUs. 
This leads to new challenges and opportunities for applying \cxlmem to the aforementioned domains. This paper aims to bridge the gap of \cxlmem understanding, and thus enable the wide adoption of \cxlmem in the community. Lastly, our \policy is specifically optimized for \cxlmem, making the most out of the memory bandwidth available for a given system.

Since the inception of the concept in 2019, 
CXL has been heavily discussed and invested by researchers and practitioners. For instance, Meta envisioned using \cxlmem for memory tiering and swapping~\cite{tpp,tmo}; Microsoft built a \cxlmem prototype system for memory disaggregation exploration~\cite{pond, 10034802}. Most of them used NUMA servers to emulate the behavior of \cxlmem.
There are also efforts in building software-based CXL simulators~\cite{yang2023cxlmemsim,wang2022asynchronous}.
Gouk \etal built a CXL memory prototype on FPGA-based RISC-V CPU~\cite{directcxl}. 
There are also a body of work focusing on certain particular applications~\cite{9969883,10032695,10066614}.
Different from the prior studies, this paper presents the first comprehensive study on true \cxlmem and compares it with emulated \cxlmem using the commercial high-performance CPU and CXL devices with both microbenchmarks and widely-used applications, which can better help the design space exploration of both CXL-memory based software systems and simulators. %This makes our research more realistic and comprehensive. 

\begin{comment}
\begin{figure}[]
    \centering
    \begin{subfigure}{0.8\linewidth}
        \centering
        \includegraphics[width=\linewidth]{figs/spec_tune_ipc_core_16.pdf}
        \caption{SPEC 507.cactuBSSN\_r}
    \end{subfigure}
    \hfill
    \begin{subfigure}{0.8\linewidth}
        \centering
        \includegraphics[width=\linewidth]{figs/spec_tune_ipc_core_17.pdf}
        \caption{SPEC 554.roms\_r}
    \end{subfigure}
    \vspace{-5pt}
    \caption{IPC changes as \policy tune}
    \label{fig:tune-improvement}
    \vspace{-12pt}
\end{figure}
\end{comment}

% !TEX root = paper.tex
\section{Conclusion}
\label{sec:concl}

%CXL memory has emerged as a standard of memory expansion and disaggregation technologies.
%However, despite active investment on CXL memory, the understanding of true CXL memory devices and their implications on system performance is falling behind.

In this paper, we have taken a first step to analyze the device-specific characteristics of true \cxlmem and compared them with NUMA-based emulations, a common practice in CXL research. % \cxlmem. 
Our analysis revealed key differences between emulated and true \cxlmem, with important performance implications. Our analysis also identified opportunities to effectively use
CXL memory as a memory bandwidth expander for memory-bandwidth-intensive applications, which leads to the development of a CXL-memory-aware dynamic page allocation policy and demonstrated its efficacy. 
%We hope that our work can facilitate the CXL-memory ecosystem and community. We will make \arch{} and \policy{} publicly available.

% !TEX root = paper.tex
\begin{acks}
We would like to thank Robert Blankenship, Miao Cai, Bhushan Chitlur, Pekon Gupta, David Koufaty, Chidamber Kulkami, Henry Peng, Andy Rudoff, Deshanand Singh, and Alexander Yu.
This work was supported in part by grants from Samsung Electronics, PRISM, one of the seven centers in JUMP 2.0, a Semiconductor Research Corporation (SRC) program sponsored by DARPA, and NRF funded by the Korean Government MSIT (NRF-2018R1A5A1059921). Nam Sung Kim has a financial interest in Samsung Electronics and NeuroRealityVision.
\end{acks}
% \input{11-appendix}

%%%%%%%%% -- BIB STYLE AND FILE -- %%%%%%%%
\bibliographystyle{ACM-Reference-Format}
\bibliography{refs}
%%%%%%%%%%%%%%%%%%%%%%%%%%%%%%%%%%%%

\end{document}